\documentclass{pasj01}
\Received{$\langle$reception date$\rangle$}
\Accepted{$\langle$acception date$\rangle$}
\Published{$\langle$publication date$\rangle$}
\usepackage{subfigure}
\usepackage{multirow}
\usepackage{lscape}
\usepackage{times}
\usepackage{comment}
\usepackage{natbib}

\begin{document}
\SetRunningHead{H.~Yamasaki \etal}{}
%\Received{}%{yyyy/mm/dd}
%\Accepted{}%{yyyy/mm/dd}
%\Published{}%{yyyy/mm/dd}
\title{Origin of the Characteristic X-ray Spectral Variations of IRAS 13224$-$3809}
%%% begin:list of authors
% Do NOT capitalize all letters in "textsc".
\author{Hiroki \textsc{Yamasaki} \altaffilmark{1,2},
Misaki \textsc{Mizumoto} \altaffilmark{1,2},
Ken \textsc{Ebisawa} \altaffilmark{1,2},
and
Hiroaki \textsc{Sameshima}\altaffilmark{3}
 }
\altaffiltext{1}{Institute of Space and Astronautical Science, Japan Aerospace 
Exploration Agency,
3-1-1 Yoshinodai, Chuo-ku, Sagamihara, Kanagawa 252-5210}
\email{mizumoto@astro.isas.jaxa.jp}
\altaffiltext{2}{Department of Astronomy, Graduate School of Science, The University of Tokyo, 7-3-1 Hongo, Bunkyo-ku, Tokyo 113-0033}
\altaffiltext{3}{Koyama Astronomical Observatory, Kyoto Sangyo University, Motoyama, Kamigamo, Kita-ku, Kyoto 603-8555}

%% `\KeyWords{}' always has to be placed before `\maketitle'.
 %Do NOT move this preamble from here!

\maketitle

\begin{abstract}
The Narrow-line Seyfert 1 galaxy (NLS1) IRAS 13224$-$3809 is known to exhibit significant X-ray spectral variation,  
a sharp spectral drop at $\sim$ 7 keV, strong soft excess emission, and a hint of iron L-edge feature, 
which is very similar to the NLS1 1H 0707$-$495.
We have proposed the ``Variable Double Partial Covering (VDPC) model'' to explain the energy spectra and spectral variability of
1H 0707$-$495 (Mizumoto, Ebisawa and Sameshima 2014, PASJ, 66, 122).
In this model, the observed flux/spectral variations below 10 keV
within a $\sim$day are primarily caused
by change of the partial covering fraction of patchy clouds composed by double absorption layers in the line of sight.
In this paper,  we apply the VDPC model to IRAS 13224$-$3809.
Consequently, we have found that the VDPC model can explain the observed spectral variations of IRAS 13224--3809 in the 0.5--10~keV band.
In particular, we can explain the observed 
 Root Mean Square (RMS) spectra (energy dependence of the fractional flux variation) in the entire 0.5 --10 keV
band. 
In addition to the well-known significant drop in the iron K-band,  we have found 
intriguing iron L-peaks in the RMS spectra
when the iron L-edge is particularly deep.
This feature, which is also found in 1H 0707--495, is naturally explained with the VDPC model,
such that the RMS variations increase at the energies where optical depths of the partial absorbers are large.
The absorbers have a larger optical depth at the iron L-edge than
in the adjacent energy bands,
thus a characteristic iron L-peak appears.
On the other hand, just below the iron K-edge,
the optical depth is the lowest and the RMS spectrum has a broad dip.
\end{abstract}

%%%%%%%%%%%%%%%%%%%%%%%%%%%%%%%%%%%%%%%

\section{Introduction}
Among Active Galactic Nuclei (AGNs), Narrow-line Seyfert 1 galaxies (NLS1s) are characterized by their particular X-ray spectral and timing properties.
A strong soft excess below $\sim$2 keV and remarkable X-ray variations are often observed,
and high-energy spectral drops at $\sim$7 keV and seemingly broadened and skewed iron emission lines are found in several objects (e.g., \citealt{boller03}).
These spectra are often explained by either the ``relativistic disk-line" model or the ``partial covering" model.
According to the ``relativistic disk-line" model, their spectra may be  interpreted by relativistically blurred inner-disk reflection around extreme Kerr black holes (e.g., \citealt{fabi04}).
On the other hand, the ``partial covering" model may also explain their spectra as due to partial covering of the central X-ray source by intervening absorbers in the line of sight (e.g., \citealt{matsu90}; \citealt{ino03}; \citealt{millerl08}). 
Furthermore, \citet{nod11,nod13} suggests that the spectral continuum of AGNs may be more complex than previously considered.
From the static spectral aspect alone, we cannot judge which model is more reasonable.
To scrutinize the validity of these models in more detail, we need to explain not only the static spectral features but also their spectral variations.  
   
NLS1s are also characterized by significant X-ray time variation.
In particular, their Root Mean Square (RMS) spectra (energy dependence of the fractional variation) tend to drop at the iron line energy band, 
which is most remarkably observed in the NLS1 MCG$-$6$-$30$-$15\footnote{Some authors argue that MCG$-$6$-$30$-$15 is a Seyfert 1 galaxy, but we treat it as NLS1 because it satisfies the properties of NLS1 \citep{Mchardy05}. } (\citealt{fabi02}; \citealt{matsumo03}). 
The ``relativistic disk-line" model explains the rapid spectral variations primarily by changes of the geometry in the very vicinity of the black hole, such as height of the illuminating source above the black hole (e.g., \citealt{mini04}).
In this model, the disk-reflected photons are much less variable than the direct photons due to relativistic reverberation (e.g., \citealt{fabi03}), thus
the characteristic RMS spectra are explained.
In addition, \cite{fabi09} reported soft lags from the NLS1 1H 0707$-$495 and interpreted them as due to reverberation from the accretion disk, where the reflection component responds to variation in the X-ray corona with the corresponding time-lag.
To the contrary, \cite{mil10} proposed that the soft X-ray lags of 1H 0707$-$495 can be accounted for by reverberation due to much more distant matter.
Up to now, similar soft lags are detected in a number of NLS1s (e.g., \citealt{kara15}),
but their origins are not fully understood.

Meanwhile, 
\citet*{mizu14} (hereafter, Paper I) successfully explained the rapid variation of 1H 0707$-$495  by the ``variable double partial covering" (VDPC) model in the 0.5--10 keV.
In this model, intrinsic X-ray luminosity and spectral shape of the central X-ray source below $\sim10$~keV are not significantly variable in timescales less than $\sim$day, 
and apparent X-ray variation is primarily caused by variation of the partial covering fraction by intervening absorbers composed of two different ionization layers.
Spectral variations in $\sim2-10$ keV and the RMS spectra of 21 Seyfert galaxies including MCG$-$6$-$30$-$15 observed with Suzaku are also explained by the VDPC model successfully (\citealt{miya12,iso16}). 
We aim to examine whether the VDPC model can explain spectral variations of other NLS1s in wider energy ranges.

In this paper, we apply the VDPC model to IRAS 13224$-$3809, which is characterized by the soft lag, significant time variation, a sharp spectral drop at $\sim$ 7 keV, strong soft excess emission, and a hint of an iron L-edge feature, being very similar to 1H 0707$-$495 (e.g., \citealt{gallo04}; \citealt{fabi13}).

%%%%%%%%%%%%%%%%%%%%%%%%%%%%%%%%%%%%%%%%

\section{Observation and Data Reduction}
We use all the available XMM-Newton \citep{jan01} and Suzaku \citep{mit07} archival data of IRAS 13224$-$3809 taken from 2001 to 2011. 
These observation IDs, observation dates, exposure times and observational modes are shown in Table \ref{tab1}.
In the following, an "observation sequence" corresponds to a row in Table \ref{tab1}.
%%%%%%%%%%%%%%%%%%%%%%%%%%%%%%%%%%%%%%%%%%%%%%%%%%%%%%%%%%%%%%%%%%%%%%%%%%%%%%%

In the analysis of the XMM-Newton data, we use the data from the European Photon Imaging Camera(EPIC)-PN \citep{str01} in 0.5$-$10 keV and the reflection grating spectrometer (RGS: \citealt{den01}) in 0.4--1.5 keV.
The data are reduced using the XMM-Newton Software Analysis System ({\tt SAS}, v.13.5.0) and the latest calibration files, following the Users guide\footnote{http://xmm.esac.esa.int/external/xmm\_user\_support/documentation/sas\_usg/USG/}. 
The event files are filtered with the conditions {\tt PATTERN<=4} and {\tt FLAG==0}. 
High background intervals when the count rates in the 10$-$12 keV band with {\tt PATTERN==0} are higher than 0.4 cts/s, are excluded.
The source spectra and light-curves are extracted from circular regions of a radius  of 35$^{\prime\prime}$ centered on the source. 
The background is made from the outer region in the same CCD chip not containing the source signals and avoiding the CCD edges.
The background subtracted light-curves are generated  with the task {\tt epiclccorr}. We apply {\tt applyabsolutecorr=yes} to the sources.
The RGS data were processed with {\tt rgsproc}.
 
%%%%%%%%%%%%%%%%%%%%%%%%%%%%%%%%%%%%%%%%%%%%%%%%%%%%%%%%%%%%%%%%%%%%%%%%%%%%%%%
In the analysis of the Suzaku data, we focus on two front-illuminated CCDs (XIS0 and XIS3) data of all the observations in 0.5$-$10 keV. 
We did not use the hard X-ray detector (HXD) PIN diode data because the source signals were hardly detected above 10 keV.
We reduce the Suzaku data by using the HEASoft version 6.16. 
As for the XIS, we screen the data with {\tt XSELECT} using the standard criterion \citep{koya07}. 
The source events are extracted from circular regions of a radius of 3$^{\prime}$  centered on the sources. The background events are extracted from an annulus of 4$^{\prime}-7^{\prime}$ in radii to avoid source regions. 
The response matrices and ancillary response files are generated for each XIS using {\tt xisrmfgen} and {\tt xissimarfgen} \citep{ishi07}.   
When we use the ARF generator, we select the number of input photons as 400,000 with the ``estepfile'' parameter ``full''.
The two XIS FI spectra and responses are combined by {\tt addascaspec}.

In Section 4.2, we will show the Suzaku observation results of Ark 564 to compare with IRAS 13224$-$3809.
Its observation ID, start date, exposure time and observational mode are shown in Table \ref{ark564_log}.
The data reduction procedure is the same as above.
%%%%%%%%%%%%%%%%%%%%%%%%%%%%%%%%%%%%%%%%%%%%%%%%%%%%%%%%%%%%%%%%%%%%%%%%%%%%%%%

%%%%%%%%%%%%%%%%%%%%%%%%%%%%%%%%%%%%%%%%%%%%%%%%%%%%%%%%%%%%%%%%%%%%%%%%%%%%%%%

\section{Results}
Paper I successfully explained spectral variations of 1H 0707$-$495 at various timescales with the VDPC model.
In this paper, we try to explain those of IRAS 13224$-$3809, following the procedure in Paper I.
In addition, we try to explain the RMS spectra in 0.5$-$10 keV with the VDPC model.
We use the X-ray spectral fitting package {\tt xspec} version 12.8.1 for spectral analysis. In the following, the {\tt xspec} model names used in the spectral fitting are indicated with the courier fonts. 
\subsection{Spectral Models}
According to Paper I, the VDPC model is expressed as 
\begin{eqnarray}
\label{eq1}
F=A_{I}(1-\alpha+\alpha W_{n} )(1-\alpha+\alpha W_{k})(P+B),
\end{eqnarray} 
where $P$ is the power-law spectrum, $B$ is the spectrum from an accretion disk ({\tt diskbb}),
$\alpha$ is the partial covering fraction,  
$A_{I}$ is the effect of interstellar absorption ({\tt phabs}),
and $W_{n}$ and $W_{k}$ are the thinner/hotter partial absorber and the thicker/colder partial absorber, respectively.
A remarkable character of the VDPC model is that the two partial absorbers have the same partial covering fraction.
Each absorber has two parameters; the hydrogen column density $N_\mathrm{H}$, and the ionization parameter $\xi$, such that $W_{i}=\exp(-\sigma (E,\xi_{i})N_{\mathrm{H},i})$, where $\sigma (E,\xi)$ is the photo-absorption cross-section.
In order to model the warm absorber, we use a table-grid model calculated with XSTAR \citep{kal04}, which is the same as the one in \citet{miya12}.
The Fe L- and K-shell edges in the observed energy spectrum are mostly explained by $W_{n}$ and  $W_{k}$, while the observed L-edge is found to be deeper than that predicted by the model from time to time.  
Also, we found absorption line features at $\sim$8 keV,
which is likely to be due to a strong Cu emission line in the outer background region (\S\ref{sec3.2}).
Consequently, the model we use to fit IRAS 13224--3809 spectra is
\begin{eqnarray}
\label{eq2}
F=A_{I}(1-\alpha+\alpha W_{n} e^{-\tau_{1}})(1-\alpha+\alpha W_{k})(P+B)+G,
\end{eqnarray} 
where $e^{-\tau_{1}}$ is the additional edge component to account for the strong Fe L-edge, 
and $G$ is a negative Gaussian.

%%%%%%%%%%%%%%%%%%%%%%%%%%%%%%%%%%%%%%%%%%%%%%%%%%%%%%%%%%%%%%%%%%%%%%%%%%%%%%%
\subsection{Spectral Fitting to the Average Spectra}\label{sec3.2}
First, we try to fit the VDPC model given in Equation \ref{eq2} to the time-average spectra for the individual observations in Table \ref{tab1}.
The interstellar absorption is fixed at the foreground absorption value from the Leiden-Argentice-Bonn 21 cm survey \citep{kalb05}.
Errors are quoted at the statistic 90\% level  throughout the paper. 

The fitting results are shown in Figure \ref{fig1} and Table \ref{IRAS_para1}.
For all the spectra, the model fit is reasonable ($\chi_{r} < 1.38$). 
Complex iron L- and K-features at $\sim 1.0$ keV and  $\sim 7.0$ keV are mostly explained by the partial covering of the thinner/hotter absorber ($W_{n}$) and the thicker/colder absorber ($W_{k}$), respectively, as expected.
The iron L-edge feature is also seen in the RGS spectra.
Figure \ref{xmm1_RGS} shows the RGS spectra of XMM1 data.
When we fit the spectra with a power-law component, we can see a sharp drop at $\sim 1.0$~keV.
When we adopt the best-fit model of the EPIC data,
we can explain the spectral feature.
This situation is similar throughout all the XMM observations.

Strong negative Gaussians at $\sim 8.0$~keV are needed in all the spectra.
There exists a strong Cu K$\alpha$ background line,
of which strength is dependent on position of the CCD.
If we over-subtract the background spectra,
absorption line features may be seen.
Figure \ref{xmm1_bkg} shows the spectra of the source/background region (green/red) and the background-subtracted spectrum (black) of XMM1 data,
which suggests that the observed negative Gaussian is an artifact.
This situation is similar in all the XMM observations.
In fact, we found the background Cu line is stronger in outer CCD area than in the central area by
visually inspecting the image created only using the Cu line band.

\subsection{Spectral Fitting of the Intensity-sliced Spectra}
Next, we apply the VDPC model to the ``intensity-sliced spectra'' within each observation sequence 
in order to investigate spectral variations.
The method to create the intensity-sliced energy spectra is as follows: (1) We create light-curves with a bin-width of 512 sec in the 0.5$-$10 keV band for each observation sequence (see Figure \ref{fig4}). 
(2) We define the four intensity ranges that contain almost equal counts (the horizontal lines in Figure \ref{fig4}). 
(3) From the time-periods corresponding to the four intensity ranges, we create four intensity-sliced energy spectra for each observation sequence.

At first, we fitted the intensity-sliced spectra within each observation sequence, varying not only the partial covering fraction $\alpha$ but also the normalizations of $P$ and $B$.
As a result, we found little variations in the normalizations of $P$ and $B$, whereas $\alpha$ varies significantly.
This situation is the same as in Paper I, where the spectral variation below $\sim$10 keV is mostly explained by change of $\alpha$ within a timescale below a $\sim$day.
Thus, we also made an assumption that the intrinsic source luminosity and the spectral shape are invariable within a $\sim$day below 10 keV, and that
only change of the partial covering fraction causes the apparent spectral variation in this energy band.
Figure \ref{IRAS_slice} and Table \ref{IRAS_slice_tab} show the fitting results,
where we can fit the four variable spectra in 0.5--10~keV only changing the covering fraction.
We also find that parameters of the disk blackbody component and the power-law component significantly differ depending on different observation sequences, indicating that the intrinsic luminosity and spectra are variable in timescales longer than $\sim$ days.

We emphasize that the spectral shape is certainly variable.
Figure \ref{alphafreeze1} shows the fitting result of XMM1 when the normalization is variable whereas the covering fraction is fixed.
We notice significant residuals in the 1--2~keV band,
which are found in all the other observation sequences.
The $\chi_{r}$ is 1.84 (d.o.f.$=688$), which is much worse than that in Table \ref{IRAS_slice_tab}.

%%%%%%%%%%%%%%%%%%%%%%%%%%%%%%%%%%%%%%%%%%%%%%%%%%%%%%%%%%%%%%%%%%%%%%%%%%%%%%%

\subsection{Energy Dependent Light Curves}
In the precedent subsection, the intensity-sliced spectra in 0.5--10 keV are explained by only change of the covering fraction, 
where the luminosity is invariable within a $\sim$ day. 
Next, following Paper I, we try to explain shorter timescale variations with the VDPC model.
To that end, we create simulated light-curves using the VDPC model for several different energy bands, and compare the model light-curves with the observed ones.

The method to calculate the simulated light-curve is as follows: 
(1) For each observation sequence, we fix all the model parameters but the partial covering fraction at the best-fit values obtained from the intensity-sliced spectral analysis in Section 3.3.
(2) For each light-curve bin, which is 512 sec long, the partial covering fraction value is calculated so that the observed counting rate in 0.5$-$10 keV and the model counting rate match.
(3) Given the partial covering fraction value thus determined for each light-curve bin, we create the simulated spectrum using {\tt fakeit} command in xspec, and calculate the simulated count-rates in 0.5$-$1.0 keV (soft), 1.0$-$3.0 keV (medium), and 3.0$-$10 keV (hard) respectively.
(4) We compare the simulated light-curves in the three bands with the observed ones.

In Figure \ref{IRAS-fig4}, we compare the observed light-curves with the model light-curves.
For all the observation sequences, the VDPC model almost perfectly reproduces the observed light-curves in the soft band,
whereas the agreement goes less satisfactory toward higher energy bands; this situation is exactly the same as 1H 0707--495.
In the hard band, the spectral variation is not fully described 
by only change of the partial covering fraction, and 
the residual is considered to be intrinsic variation in the hard energy band.
This situation seems to be consistent with the assumption in the ``optxagn'' model \citep{don12},
where the hard power-law component is produced by variable hot/thin coronal emission, 
whereas the soft component is associated with the warm/thick Comptonizing layer on the disk surface that up-scatters the intrinsic disk emission.
In this scenario, 
the hard power-law component is more variable than the soft component.

%%%%%%%%%%%%%%%%%%%%%%%%%%%%%%%%%%%%%%%%%%%%%%%%%%%%%%%%%%%%%%%%%%%%%%%%%%%%%%%%
\subsection{The Root-mean-square (RMS) spectra}
Next, we calculate the RMS spectra from the VDPC model to compare with the observed ones.
\citet{miya12} and \citet{iso16} successfully explained the RMS spectra of MCG$-$6$-$30$-$15 and other $\sim$20 Seyfert galaxies in 2--10 keV by the VDPC model. 
In particular, the characteristic iron Fe-K feature, significant drop of the fractional variation at the broad iron line energy,  was accounted for.
Here, we also calculate the RMS spectra including not only the Fe-K band but also the Fe-L band.
The method to calculate the RMS spectra is as follows;
(1) Following the analysis in section 3.3 and 3.4, we create the observed and simulated light-curves in 16 energy bands.
(2) Following \citet{edel02}, RMS variations from the observed light-curves are calculated with a time-bin width of $2.3\times 10^{4}$ sec\footnote{Corresponding to four times the Suzaku orbital period. Following \citet{iso16}, we choose this time-bin width to minimize the influence of the discontinuity of data.} in 0.5$-$10 keV in each of the 16 energy bands.
(3) The simulated RMS spectra are calculated from the simulated light-curves to compare with the observed ones.

In Figure \ref{IRAS_rms_spe}, we show the observed RMS spectra and the model RMS spectra for each observation. The observed RMS spectra are explained by the VDPC model. 
In addition to the well-known Fe-K feature,
we find characteristic peaks at around 1.1 keV, the iron L-edge energy band. This feature is most clearly seen when the iron L-edge is particularly deep in the intensity-sliced spectra (XMM1 in Figure \ref{IRAS_slice} ).

\section{Discussion}
\subsection{Interpretation of the spectral variations}\label{sec:4.1}
In Section 3.4, we aimed to explain the observed
0.5--10 keV light-curves in a timescale shorter than a $\sim$ day with only change of the partial covering fraction by the VDPC model.
As a result, the observed light-curves below 3 keV can be explained by the VDPC model.
Meanwhile, above 3 keV, the observed light-curves tend to show slight deviations from the simulated ones, and the difference is greater in the high energy band.
This result is exactly the same as in the case of  1H 0707--495 (Paper I), suggesting that intrinsic variation of the hard spectral component ($P$ in Equation \ref{eq2}) is not negligible above 3 keV. Consequently,  the observed  X-ray spectra variation is explained  presumably  by two independent variations of physical parameters; the partial covering fraction,
which accounts for most of the soft X-ray spectral variations,
and the intrinsic variation of the hard spectral component, which is more significant above $\sim3$~keV.

In order to study both variations of the partial covering fractions and the hard spectral component,
we need to analyze the data in wider energy band.
In the hard X-ray energy band ($>20$~keV), effect of the partial covering is negligible, thus
the observed flux variation shall represent the intrinsic luminosity variation of the hard component.
In fact, simultaneous observations of NuSTAR and XMM-Newton of MCG$-$6$-$30$-$15 suggests that the apparently complicated  spectral variations in 0.2$-$60 keV in timescales less than $\sim$day
are naturally explained by independent variations of the partial covering fraction and normalization of the hard spectral component (Kusunoki et al.\ in prep.).

We also point out that the observed apparently strong iron spectral features can be explained with a solar abundance absorber in our model, 
whereas some papers have argued
extreme iron-overabundance (more than 10 times) in this AGN (see, e.g.\,\citealt{fabi13,chi15}).
This is because the particular spectral shape of the VDPC model (eq.\ref{eq2}) has strong spectral troughs at iron L- and K-edges even with the solar abundance.

\subsection{Interpretation of the Root-mean-square Spectra}
In the RMS spectra of IRAS 13224$-$3809 (Figure \ref{IRAS_rms_spe}), we have found characteristic peaks at $\sim1.1$~keV.
Similar peaks at $\sim1.1$~keV are also recognized in Figure 3 of \citet{pon10} and Figure 11 of \citet{fabi13} on the same target.
The iron L-feature must be responsible for this RMS peak because they are in the same energy range.
In order to investigate effect of the iron L-edge to the RMS spectra,
we create RMS spectra of 1H 0707--495, which has a strong iron L-edge, and Ark 564, which has hardly iron L-feature, in their energy spectra.
As a result, we find that 1H 0707--495 also has an iron L-peak in the RMS spectra, whereas that of Ark 564 show no structure around iron L-energy band at all (Figure \ref{1H0707}).
A similar RMS peak is also seen in NGC 5548, which has a sharp iron L-edge in the energy spectrum \citep{cap16}.
Thus, the characteristic RMS peak is certainly originated from the iron L-edge.
In addition, broad dips are seen at $\sim7$~keV both in the RMS spectra of IRAS 13224--3809 and 1H 0707--495, which are due to iron K-structure (see, e.g.\,Figure 9 of \citealt{matsumo03}).

In Figure \ref{IRAS_rms_spe} and \ref{1H0707}, the RMS spectra in 0.5--10 keV are successfully explained by the VDPC model.
What makes the characteristic structures of the RMS spectra, the iron L-peak and K-dip, in the VDPC model?
In the VDPC model, spectral variation below $\sim10$~keV is primarily caused by change of the partial covering fraction.
When the intrinsic source luminosity and the spectrum are not variable, 
the observed spectral variation, $F_{obs}(E, t)$, due to variable covering fraction, $\alpha(t)$, may be expressed as (see eq.\,\ref{eq1}),  
\begin{eqnarray}
{\scriptstyle F_{obs}(E, t)} &\propto& {\scriptstyle  \left(1 - \alpha(t) + \alpha(t) \; e^{-\tau_n(E)} \right) \left(1 - \alpha(t) + \alpha(t) \; e^{-\tau_k(E)}\right) \label{eqa}} \\
& = & {\scriptstyle \left(1 - \alpha(t) (1 -  e^{-\tau_n(E)}) \right) \left(1 - \alpha(t) (1-   \; e^{-\tau_k(E)})\right) \label{eqb}} \\
& \approx &{\scriptstyle \left(1 - \alpha(t)\: \tau_n(E) \right) \left(1 - \alpha(t) (1-\; e^{-\tau_k(E)})\right)\;   {\rm when}\;  \tau_n(E) \ll 1 \label{eqc}} \\
& \approx & {\scriptstyle \left(1 - \alpha(t) (1 - e^{-\tau_n(E)}) \right) (1 - \alpha(t)) \;   {\rm when}\;    \tau_k(E) \gg 1,  \label{eqd} }
\end{eqnarray}
where $\tau_n(E)$ and $\tau_k(E)$ are optical depths of the thinner (hotter) and thicker (colder) absorbers, 
which primarily responsible for the observed iron L-edge and K-edge, respectively.
The upper panel of Figure \ref{RMSsimulation} shows the model spectra where the covering fraction is variable from 0.01 to 0.99.

Let's consider energy dependence of the spectral variation  due to  variation of $\alpha(t)$.
From Eq.~(\ref{eqb}), we see that the spectral variation tends to be larger at the energies where absorbers are optically thick.
At around the iron K-band, where $\tau_n(E) \ll 1$, the spectral variation is represented as Eq.~(\ref{eqc}).
From lower energies toward the iron K-edge energy, $\tau_n(E)$ continuously decreases, and at the iron K-edge,  $\tau_k(E)$  suddenly increases.  Consequently, the 
spectral variation will be minimum just before the iron K-edge,
where the broad trough appears in the simulated RMS spectrum (Figure \ref{RMSsimulation} bottom).
To the contrary, in lower energy range where $\tau_k(E) \gg 1$, the spectral variation is represented as Eq.~(\ref{eqd}).
At the iron-L edge, where $\tau_n(E)$ is the largest,
spectral variation is most significant; thus 
a characteristic broad peak appears in the RMS spectrum at around iron L-edge (Figure \ref{RMSsimulation} bottom).

Also, we point out that many peaks are expected in the RMS spectrum corresponding to absorption lines in the energy spectrum (Figure \ref{RMSsimulation} bottom).
If absorbers are static, we do not see such peaks in the RMS spectra. 
In this manner, by studying RMS variation of individual absorption lines, 
we may distinguish multiple absorption layers having different variation timescales.
This method can be applicable to the data with higher energy resolution such as those taken by RGS on XMM, or, more effectively, by future microcalorimeter instruments (Mizumoto \& Ebisawa, in prep.).

\section{Conclusion}
We have studied spectral variations of NLS1 IRAS 13224$-$3809, using all the currently available XMM-Newton and Suzaku archival data.
Following Paper I, we examined if the observed spectral variation is explained by the Variable Double Partial Covering (VDPC) model. Consequently, we have found that the VDPC model can successfully explain the averaged and intensity-sliced spectra of IRAS 13224$-$3809 in 0.5$-$10 keV within a $\sim$ day only changing the partial covering fraction. 
The model can explain the light-curves within a $\sim$day
mostly by only change of the partial covering fraction,
whereas some intrinsic variation above $\sim$ 3 keV is additionally recognized.
We have successfully explained the observed RMS spectra in the entire 0.5$-$10 keV band with the VDPC model.
In addition to the well-known significant drop in the iron K-band, 
we have found such intriguing broad iron L-peaks in the RMS spectra (as well as 1H 0707$-$495), that is particularly significant when the iron L absorption edge is deep in the energy spectra. 
These RMS spectral features can be explained by only change of the partial covering fraction,
such that the RMS variation increases at the energies where the optical depth of the partial absorbers is large, and vice versa.
The optical depth is minimum just below the iron K-edge and suddenly increases at the iron K-edge, 
thus the broad dip structure is produced.
Around the iron L-energy band, the optical depth is the largest,
thus the characteristic peak appears.

\bigskip
\bigskip
This research has made use of public Suzaku data obtained through the Data ARchives
and Transmission System (DARTS), provided by Institute of Space and Astronautical
Science (ISAS) at Japan Aerospace Exploration Agency (JAXA). This work is also based
on observations obtained with XMM-Newton, an ESA science mission with instruments and
contributions directly funded by ESA Member States and the USA (NASA), and the XMM-Newton
data obtained through the XMM-Newton Science Archive at ESA.
For data reduction, we used software provided by the High Energy Astrophysics Science Archive Research Center at NASA/Goddard Space Flight Center.
MM and KE are financially supported by the Japan Society for the Promotion of Science (JSPS) KAKENHI Grant Number 15J07567 and 16K05309, respectively. 
We acknowledge the referee, Dr.~J.~Reeves, for valuable comments.
%%%%%%%%%%%%%%%%%%%%%%%%%%%%%%%%%%%%%%%%%%%%%%%%%%%%%%%%%%%%%%%%%%%%%%%%%%%%%%%
\bibliographystyle{aa}
\bibliography{reference}

%\end{thebibliography}

%%%%%%%%%%%%%%%%%%%%%%%%%%%%%%%%%%%%%%%%%%%%%%%%%%%%%%%%%%%%%%%%%%%%%%%%%%%%%%%

\clearpage
%IRAS log

\begin{longtable}{ccccc}
\caption{Suzaku and XMM-Newton observations of IRAS 13224$-$3809. Observation IDs, start dates, exposure times and the observation modes of XMM/PN and Suzaku/XIS are indicated. The exposure time of Suzaku is that of XIS 0.  }\\

\hline
Name     &Observation ID&Date           &Exposure  &Obs. Mode          \\ \hline
XMM1    &0110890101     &2001-01-19&60.9$\ $ks   & Full Frame         \\                 
XMM2    &0673580101     &2011-07-19&126.1$\ $ks & Full Frame    \\ 
XMM3    &0673580201     &2011-07-21&125.1$\ $ks & Large Window     \\ 
XMM4    &0673580301     &2011-07-25&125.0$\ $ks & Large Window      \\ 
XMM5    &0673580401     &2011-07-29&127.5$\ $ks & Large Window     \\ 
Suzaku1 &701003010      &2007-01-26&198.0$\ $ks & Full Window  
\\ \hline
\endhead
\hline
\label{tab1}
\endfoot
\end{longtable}

%Ark 564 
%log
\begin{longtable}{ccccc}
\caption{Suzaku observation of Ark 564. Observation ID, start date, valid exposure time and the observation mode of XIS are indicated. The exposure time is that of XIS 0. }\\

\hline
Observation ID&Date           &Exposure  &Obs. Mode          \\ \hline
%XMM1    &0206400101  &2005-01-05&99.0$\ $ks   &Small Window     \\                 
702117010    &2007-06-26&100.0$\ $ks  &Full Window \\ 
\hline
\endhead
\hline
\label{ark564_log}
\endfoot
\end{longtable}

%IRAS average  parameter

\begin{landscape}

\begin{table}

\caption{Results of the average spectral fitting for IRAS13224$-$3809.  }\label{IRAS_para1}

\begin{center}

\begin{tabular}{llllllllll}

\hline
 &         & XMM1& XMM2&XMM3&XMM4&XMM5&Suzaku1 \\ \hline

\hline

A$_{I}$&$N_{\rm H}$\ ($10^{20}\ \rm cm^{-2}$)&
5.34(fix)&
5.34(fix)&
5.34(fix)&
5.34(fix)&
5.34(fix)&
5.34(fix)\\ \hline

W$_{k}$&
$N_{\rm H}$ ($10^{23}\ \rm cm^{-2}$)&
$5.9_{-1.5}^{+2.3}$&
$>12$&
$6.3_{-1.4}^{+1.9}$&
$6.1_{-1.9}^{+2.2}$&
$7.5_{-0.7}^{+3.2}$&
$>10$\\

                        &
$\log \xi$&
$1.7^{+0.4}_{-1.1}$&
$0.36^{+0.12}_{-0.19}$&
$0.10_{-0.10}^{+0.08}$&
$0.10_{-0.10}^{+0.09}$&
$0.1\pm0.1$ &
$0.36_{-0.13}^{+0.09}$ \\ \hline

W$_{n}$&
$N_{\rm H}$ ($10^{23}\ \rm cm^{-2}$)&
$0.99^{+1.95}_{-0.39}$&
$2.8^{+6.1}_{-2.6}$&
$0.21_{-0.13}^{+0.70}$&
$3.5_{-2.6}^{+2.7}$&
$0.25_{-0.16}^{+1.55}$&
$8.8_{-4.5}^{+3.2}$\\ 

&
$\log \xi$&
$2.95^{+0.12}_{-0.05}$&
$3.6^{+0.9}_{-0.2}$&
$2.95_{-0.04}^{+0.17}$&
$3.05_{-0.21}^{+0.05}$ &
$2.95_{-0.08}^{+0.19}$&
$3.14_{-0.04}^{+0.03}$ \\

Edge                &
$E_{\rm cut}$ (keV) &
$1.09^{+0.01}_{-0.01}$&
$1.06\pm 0.01$&
$1.11\pm0.01$&
$1.067\pm 0.009$ &
$1.08\pm 0.01$&
$1.07\pm 0.01$\\                      

&
$\tau$&
$1.4^{+0.3}_{-0.2}$&
$0.43\pm 0.05$&
$0.73_{-0.09}^{+0.08}$&
$2.0_{-0.4}^{+0.7}$ &
$0.7\pm0.1$&
$1.0_{-0.3}^{+0.8}$\\ \hline

$P$&
Photon index&
$3.0\pm0.2$&
$2.65^{+0.11}_{-0.08}$&
$2.90\pm 0.07$&
$2.9\pm 0.2$ &
$2.7\pm 0.1$&
$2.5\pm 0.1$\\ 

                         &
norm$^{a}$&
$2.3^{+0.8}_{-0.6}$&
$2.6^{+0.6}_{-0.9}$&
$1.79_{-0.18}^{+0.07}$&
$1.2_{-0.4}^{+0.5}$ &
$1.3_{-0.2}^{+0.4}$&
$1.2_{-0.3}^{+0.4}$\\ \hline

$B$&
$T_{\rm in}$ (keV)&
$0.167^{+0.011}_{-0.007}$&
$0.167\pm0.003$& 
$0.162_{-0.003}^{+0.004}$ &
$0.152\pm0.004$&
$0.159_{-0.002}^{+0.003}$&
$0.160_{-0.006}^{+0.008}$ \\ 

                         &
norm$^{b}$&
$1.2 \pm 0.4 \times 10^{3}$&
$1.7^{+0.4}_{-0.6} \times 10^{3}$&
$1.11_{-0.05}^{+0.07}\times 10^{3}$&
$1.2\pm0.3\times 10^{3}$&
$1.3\pm0.2 \times 10^{3}$&
$1.0_{-0.3}^{+0.5}\times 10^{3}$  \\ \hline

$\alpha$&
&
$0.74^{+0.09}_{-0.07}$&
$0.75^{+0.03}_{-0.14}$&
$0.70_{-0.07}^{+0.04}$&
$0.73\pm 0.06$&
$0.63_{-0.07}^{+0.11}$&
$0.60_{-0.12}^{+0.09}$ \\ \hline
$G$                 &
E (keV) &
$8.45^{+0.05}_{-1.04}$&
$8.02^{+0.29}_{-0.05}$&
$7.83_{-0.05}^{+0.13}$&
$8.10\pm 0.05$ &
$8.48_{-0.05}^{+0.03}$&
$7.87_{-0.22}^{+0.15}$                               \\
                       
                         &
Sigma (keV) &
0.01(fix)&
0.01(fix)&
0.01(fix)&
0.01(fix)&
0.01(fix)&
0.01(fix)                            \\
 
                         &
Norm  ($10^{-6} \rm \ photons \ s^{-1} \ cm^{-2}$) &
$-3.3^{+1.7}_{-1.1}$&
$-1.8\pm0.8$&
$-1.1\pm 0.5$&
$-1.4\pm0.6$&
$-2.5\pm0.7$&
$-0.5\pm0.4$                                                                                       \\ \hline
    &
Reduced chisq (d.o.f)&
1.38(254)&
1.18(326)&
0.97(325)&
0.96(222)&
1.12(307)&
1.05(482) \\ \hline   

    \end{tabular}
 \\   
$^{a}$Photon flux at 1 keV in units of  $10^{-3} \rm \ photons  \ s^{-1} \ cm^{-2}$. \\
$^{b}$Diskbb normalization, $((R_{{\rm in}})/({\rm km})/(D/10 \ {\rm kpc}))^{2} \cos \theta$.    
    \end{center}

\end{table}
\end{landscape}

%IRAS slice parameter

\begin{landscape}
\begin{table}
\caption{Results of the slice spectral fitting for IRAS13224$-$3809. See the caption of Table \ref{IRAS_para1} for the explanation of parameters.}\label{IRAS_para2}
\label{IRAS_slice_tab}
\begin{center}

\begin{tabular}{llllllllll}
\hline
&                                                      &XMM1&XMM2&XMM3&XMM4&XMM5&Suzaku1 
\\ \hline

A$_{I}$&
$N_{\rm H}$\ ($10^{20}\ \rm cm^{-2}$)&
5.34(fix)&
5.34(fix)&
5.34(fix)&
5.34(fix)&
5.34(fix)&
5.34(fix)\\ \hline

W$_{k}$&
$N_{\rm H}$ ($10^{23}\ \rm cm^{-2}$)&
$10\pm2$&
$>13$&
$14.3_{-0.5}^{+3.0}$&
$14.5_{-0.5}^{+0.9}$&
$12.1_{-0.4}^{+0.8}$&
$14_{-1}^{+4}$
\\
 
                        &
$\log \xi$&
$0.1_{-0.1}^{0.4}$&
$0.1\pm0.1$&
$0.4 \pm 0.1$&
$0.36\pm0.04$&
$0.36_{-0.05}^{+0.08}$&
$0.36_{-0.09}^{+0.10}$
 \\ \hline

W$_{n}$&
$N_{\rm H}$ ($10^{23}\ \rm cm^{-2}$)              &
$8.3_{-3.7}^{+4.1}$&
$8.9_{-3.0}^{+3.6}$&
$0.5_{-0.3}^{+2.1}$&
$2.9_{-2.3}^{+1.8}$&
$11_{-1}^{+2}$&
$6.3_{-2.9}^{+3.2}$
\\
 
                        &
$\log \xi$&
$3.08_{-0.05}^{+0.03}$&
$3.13_{-0.03}^{+0.02}$&
$3.0\pm 0.1$&
$3.06_{-0.14}^{+0.04}$&
$3.11\pm 0.01$&
$3.12_{-0.03}^{+0.02}$ 
\\

Edge                &
$E_{\rm cut}$ (keV) &
$1.06 \pm 0.01$&
$1.08 \pm 0.01$&
$1.10 \pm 0.01$&
$1.071_{-0.004}^{+0.007}$&
$1.077\pm 0.006$&
$1.07\pm 0.01$
    \\                    
                        &
$\tau$&
$3.4_{-0.5}^{+0.7}$&
$1.2_{-0.1}^{+0.2}$&
$0.65\pm 0.06$&
$1.4\pm0.1$&   
$1.51_{-0.10}^{+0.09}$&
$0.9\pm 0.1$
\\ \hline

$P$       &
Photon index &
$2.7 \pm0.2 $&
$2.68_{-0.06}^{+0.08}$&
$2.58_{-0.08}^{+0.11}$&
$2.54_{-0.09}^{+0.06}$&
$2.77_{-0.03}^{+0.06}$&
$2.58_{-0.03}^{+0.07}$
\\
 
                         &
norm$^{a}$&
$1.5_{-0.3}^{+0.5}$&
$1.6\pm0.1$& 
$1.7_{-0.3}^{+0.5}$&
$1.5_{-0.2}^{+0.1}$&
$2.5_{-0.1}^{+0.2}$&
$1.58_{-0.05}^{0.08}$&
\\ \hline

$B$              &
$T_{\rm in}$ (keV)&
$0.168_{-0.004}^{+0.005}$&
$0.175\pm 0.003$&
$0.164 \pm0.003$&  
$0.151 \pm0.003$&    
$0.172_{-0.003}^{+0.002}$&
$0.156_{-0.002}^{+0.007}$&
\\
 
                         &
norm$^{b}$ &
$9_{-2}^{+1}\times 10^{2} $ &
$6.4_{-0.8}^{+0.7}\times 10^{2}$&
$1.6_{-0.2}^{+0.4}\times 10^{3}$&
$2.8\pm0.3 \times 10^{3}$&
$9.7 _{-0.5}^{+0.7}$&
$1.42_{-0.09}^{+0.26}$
 \\ \hline

$\alpha$&
 &
$0.76\pm0.02$&
$0.712_{-0.006}^{+0.007}$&
$0.86\pm 0.01$&  
$0.934_{-0.009}^{+0.004}$&
$0.871\pm 0.02$&  
$0.823_{-0.032}^{0.008}$            
\\                 
                        
                         &                                                               
                         &

$0.61\pm0.04$&   
$0.534_{-0.007}^{+0.01}$&
$0.75\pm0.04$&
$0.82_{-0.02}^{+0.01}$&
$0.701\pm0.04$&   
$0.62_{-0.06}^{+0.02}$
\\                        
                         &
                         &

$0.48\pm0.05 $&
$0.424_{-0.007}^{+0.013}$&
$0.67_{-0.05}^{+0.07}$&
$0.74_{-0.03}^{+0.01}$&
$0.465\pm0.006$&
$0.533_{-0.079}^{+0.009}$&
\\

                         &
                         &

$0.33\pm 0.06$& 
$0.25_{-0.25}^{+0.02}$&
$0.51_{-0.07}^{+0.09}$&
$0.65_{-0.04}^{+0.03}$&
$0.300_{-0.300}^{+0.003}$&
$0.38_{-0.10}^{+0.02}$
\\ \hline
    
$G$                 &
E (keV)             & 
$8.1_{-0.1}^{+1.3}$&
$7.98_{-0.07}^{+0.22}$&
$8.2\pm0.8$& 
$8.18_{-0.44}^{+0.07}$&
$8.09_{-0.06}^{+0.64}$&
$7.63_{-0.15}$
\\
                                                                                      
                         &
Sigma (keV)   &  
0.01(fix)&
0.01(fix)&
0.01(fix)&
0.01(fix)&
0.01(fix)&
0.01(fix)&
\\                                                                             
                         
                         &
Norm  (10$^{-6}$ photons s$^{-1}$ cm$^{-2}$) &
$-2.1\pm2.1$&    
$-1.8\pm0.8$&
$-1.5\pm0.6$&  
$-1.3_{-0.8}^{+0.7}$&
$-1.7_{-0.7}^{+0.6}$&   
$-0.6\pm0.6$
\\ \hline                         
              
                         &
Reduced chisq (d.o.f) &
1.20(688)&    
1.12(998)&    
1.19(927)&    
1.12(615)&    
1.31(848)&    
1.03(496)
 \\ \hline

    \end{tabular}
\\    
$^{a}$Photon flux at 1 keV in units of  $10^{-3} \rm \ photons  \ s^{-1} \ cm^{-2}$. \\
$^{b}$Diskbb normalization, $((R_{{\rm in}})/({\rm km})/(D/10 \ {\rm kpc}))^{2} \cos \theta$.      
    \end{center}

\end{table}
\end{landscape}
%%%%%%%%%%%%%%%%%%%%%%%%%%%%%%%%%%%%%%%%%%%%%%%%%%%%%%%%%%%%%%%%%%%%%%%%%%%%%%%

%IRAS average spectra

\addtocounter{figure}{0}
\begin{figure*}[ht]
  \begin{minipage}{0.5\hsize}
    \begin{center}
     \rotatebox{-90}{ \FigureFile(60mm,80mm){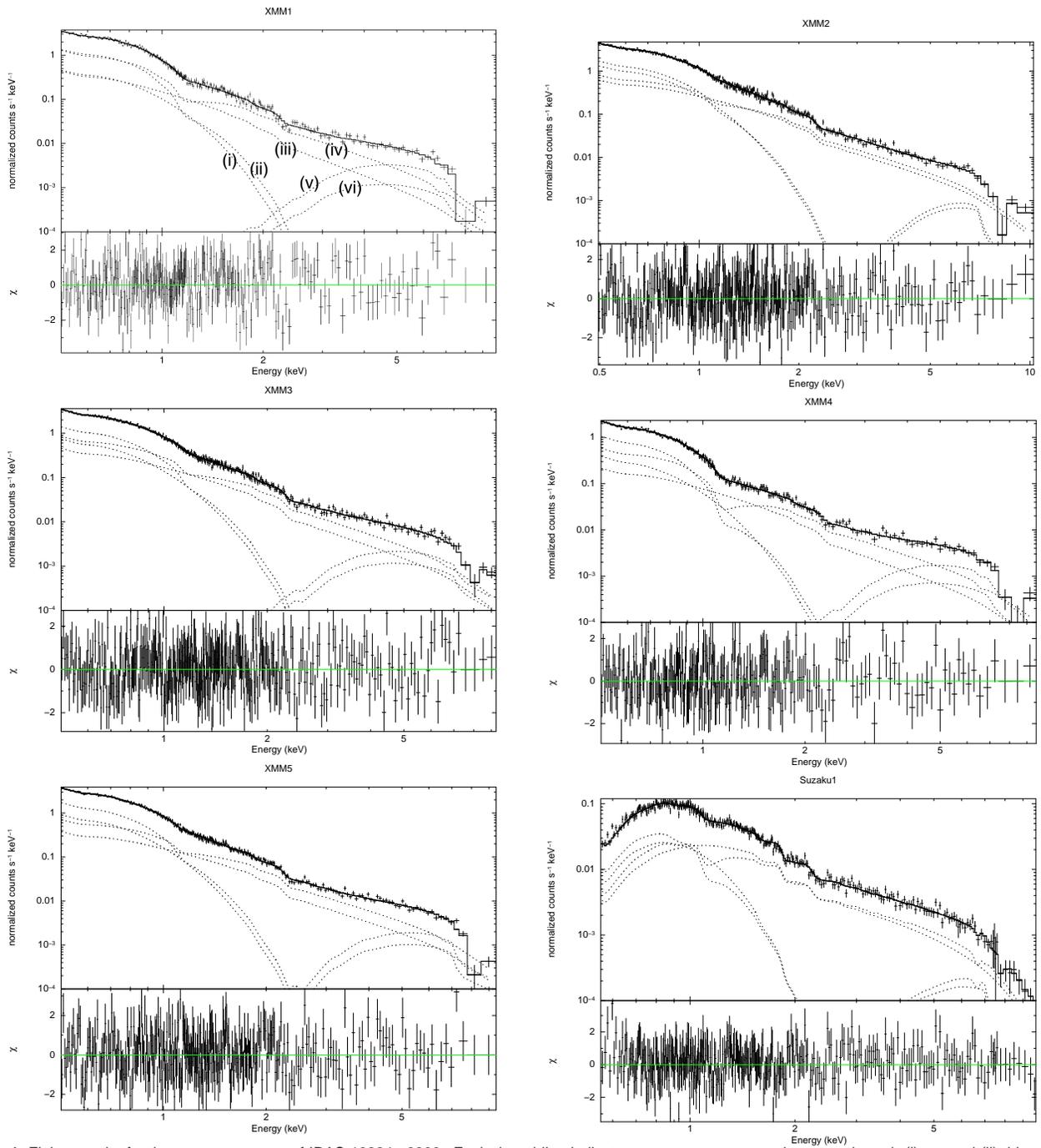}} \\
   \rotatebox{-90}{\FigureFile(60mm,80mm){f1c.eps}} \\
 \rotatebox{-90}{\FigureFile(60mm,80mm){f1e.eps}} \\
    \end{center}
  \end{minipage}
\begin{minipage}{0.5\hsize}
 \begin{center}
     \rotatebox{-90}{\FigureFile(60mm,80mm){f1b.eps}} \\
     \rotatebox{-90}{\FigureFile(60mm,80mm){f1d.eps}} \\
      \rotatebox{-90}{\FigureFile(60mm,80mm){f1f.eps}} \\
       \end{center}
  \end{minipage}
  \caption{Fitting results for the average spectra of IRAS 13224$-$3809. 
Each dotted line indicates {\tt diskbb} component that pass through (i) no, and (ii) thinner absorber, and power-law component 
that pass through (iii) no, (iv) thinner, (v) thicker, and (vi) both thinner and thicker absorbers.
The {\tt diskbb} component that passes through the thicker absorber is too weak to be seen within the panel.
}
  \label{fig1}
\end{figure*}

%%%%%%%%%%%%%%%%%%%%%%%%%%%%%%%%%%%%%%%%%%%%%%%%%%%%%%%%%%%%%%%%%%%%%%%%%%%%%%%%%%%%%

\addtocounter{figure}{0}
\begin{figure*}[ht]
\begin{center}
\rotatebox{-90}{\FigureFile(70mm,90mm){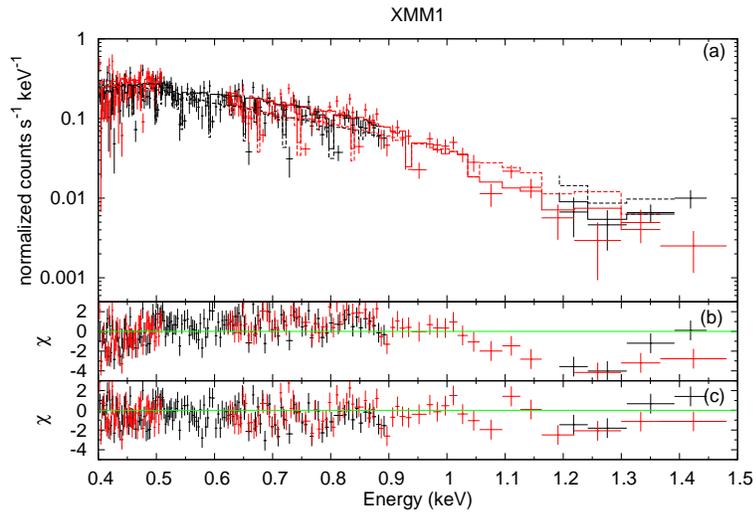}} 
\end{center}
\caption{
(a) Fitting result for the RGS 1st-order spectrum (XMM1).
The dashed line shows the power-law model, and the solid line shows the best-fit EPIC model.
(b) Residuals of the power-law model. 
(c) Residuals of the best-fit EPIC model.
}
\label{xmm1_RGS}
\end{figure*}

\addtocounter{figure}{0}
\begin{figure*}[ht]
\begin{center}
\rotatebox{-90}{\FigureFile(70mm,90mm){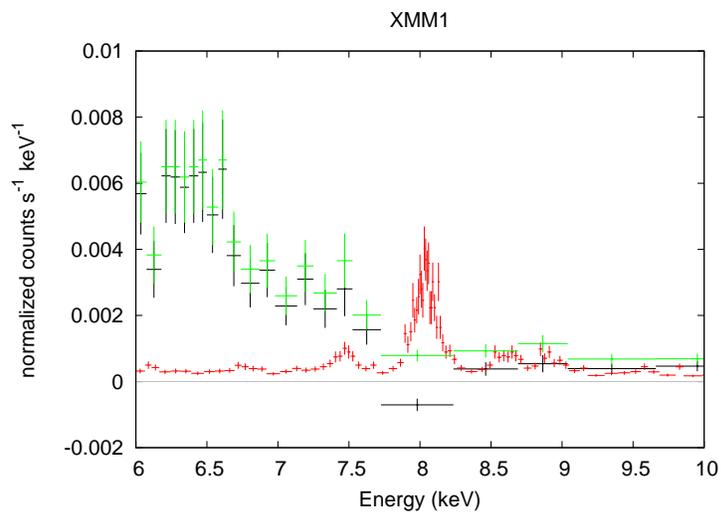}} 
\end{center}
\caption{
Spectra from the source/background region (green/red) and the background-subtracted spectrum (black) of XMM1 data.
The background spectrum is normalized according to the detector area of the source region.
}
\label{xmm1_bkg}
\end{figure*}

%IRAS lightcurve

\addtocounter{figure}{0}
\begin{figure*}[ht]
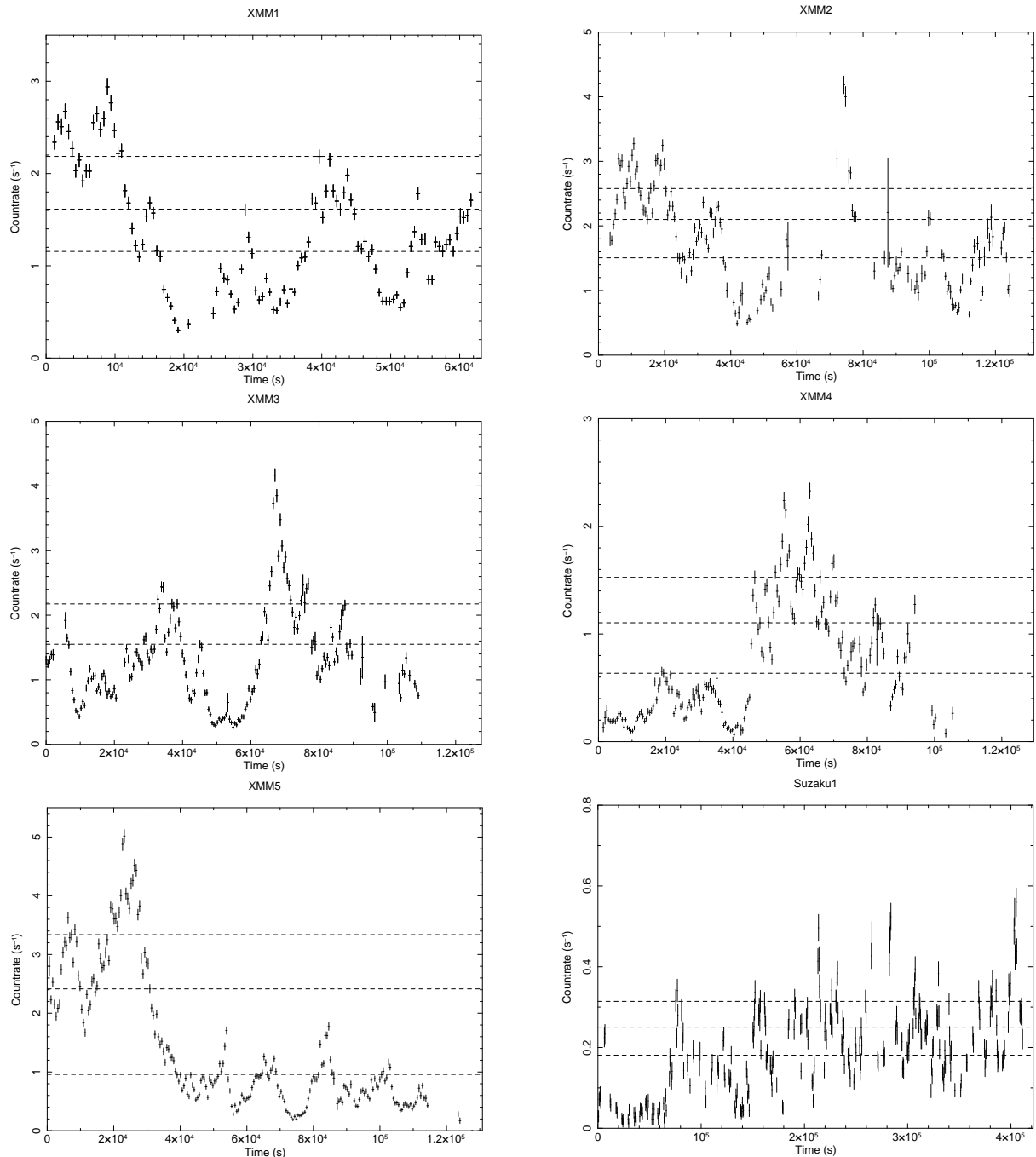

  \begin{minipage}{0.5\hsize}
    \begin{center}
     \rotatebox{-90}{ \FigureFile(60mm,80mm){f4a.eps}} \\
   \rotatebox{-90}{\FigureFile(60mm,80mm){f4c.eps}} \\
 \rotatebox{-90}{\FigureFile(60mm,80mm){f4e.eps}} \\
    \end{center}
  \end{minipage}
\begin{minipage}{0.5\hsize}
 \begin{center}
     \rotatebox{-90}{\FigureFile(60mm,80mm){f4b.eps}} \\
     \rotatebox{-90}{\FigureFile(60mm,80mm){f4d.eps}} \\
      \rotatebox{-90}{\FigureFile(60mm,80mm){f4f.eps}} \\
       \end{center}
  \end{minipage}
  \caption{Light-curves in the 0.5$-$10.0 keV band.
The horizontal dotted lines show the thresholds to create the intensity-sliced spectra.
 }
  \label{fig4}
\end{figure*}

%%%%%%%%%%%%%%%%%%%%%%%%%%%%%%%%%%%%%%%%%%%%%%%%%%%%%%%%%%%%%%%%%%%%%%%%%%%%%%%%%%%%%%

%IRAS slice 

\addtocounter{figure}{0}
\begin{figure*}[ht]
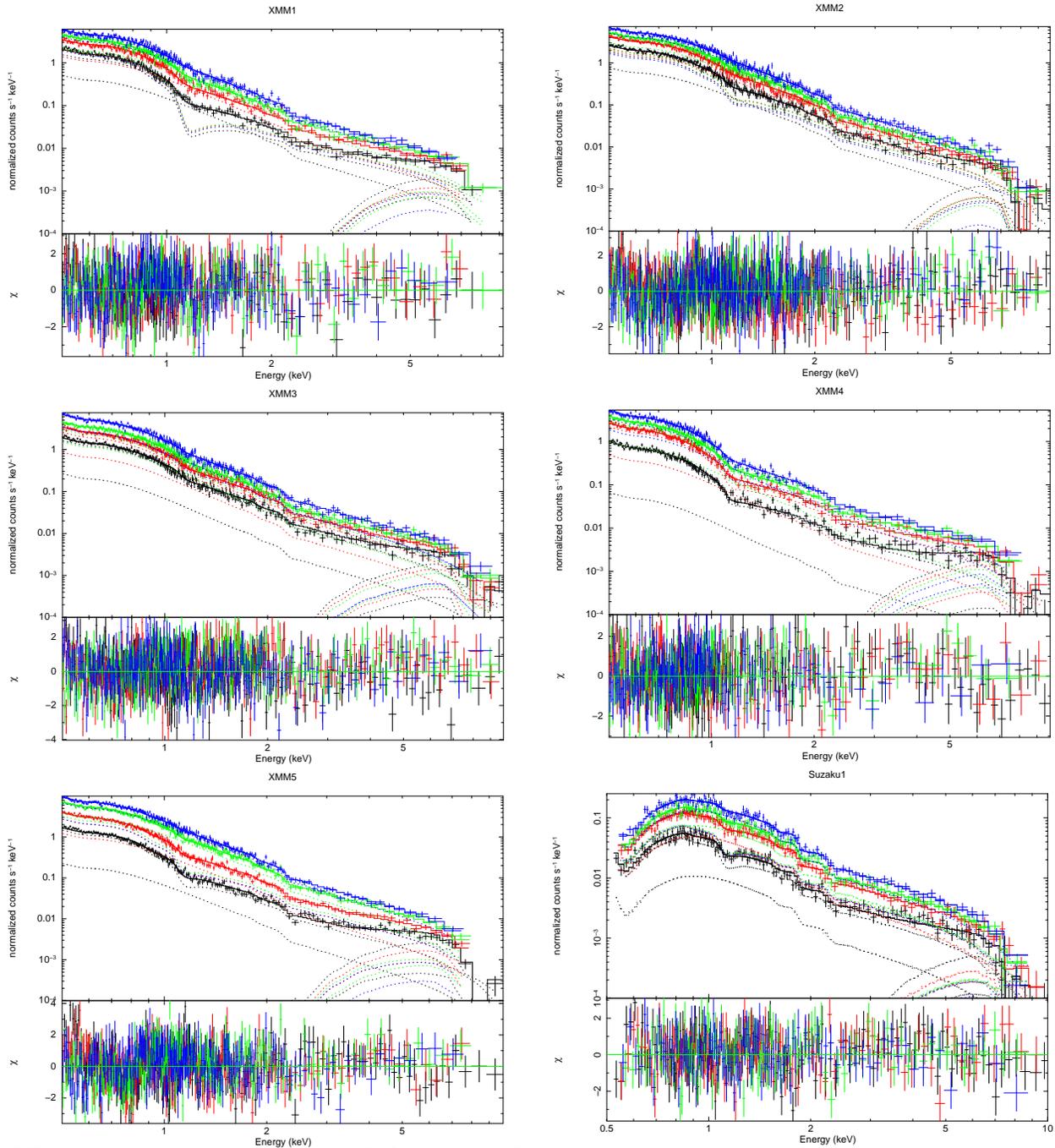

  \begin{minipage}{0.5\hsize}
    \begin{center}
     \rotatebox{-90}{ \FigureFile(60mm,80mm){f7a.eps}} \\
   \rotatebox{-90}{\FigureFile(60mm,80mm){f7c.eps}} \\
 \rotatebox{-90}{\FigureFile(60mm,80mm){f7e.eps}} \\
    \end{center}
  \end{minipage}
\begin{minipage}{0.5\hsize}
 \begin{center}
     \rotatebox{-90}{\FigureFile(60mm,80mm){f7b.eps}} \\
     \rotatebox{-90}{\FigureFile(60mm,80mm){f7d.eps}} \\
      \rotatebox{-90}{\FigureFile(60mm,80mm){f7f.eps}} \\
       \end{center}
  \end{minipage}
  \caption{Fitting results for the intensity-sliced spectra of IRAS 13224$-$3809. For each observation, the four intensity-sliced spectra are fitted simultaneously only varying the partial covering fraction.
The solid lines show the best-fit models, and the individual model components are indicated with dotted lines; the {\tt diskbb} $+$ power-law component not going through the absorbers (i + iii in Fig.\,\ref{fig1}), that through the thin absorber (ii + iv in Fig.\,\ref{fig1}), the power-law component going through the thick absorber  (v in Fig.\,\ref{fig1}) and that through the both components (vi in Fig.\,\ref{fig1}).  
}
  \label{IRAS_slice}
\end{figure*}

%%%%%%%%%%%%%%%%%%%%%%%%%%%%%%%%%%%%%%%%%%%%%%%%%%%%%%%%%%%%%%%%%%%%%%%%%%%%%%%%%%%%%%

\addtocounter{figure}{0}
\begin{figure*}[ht]
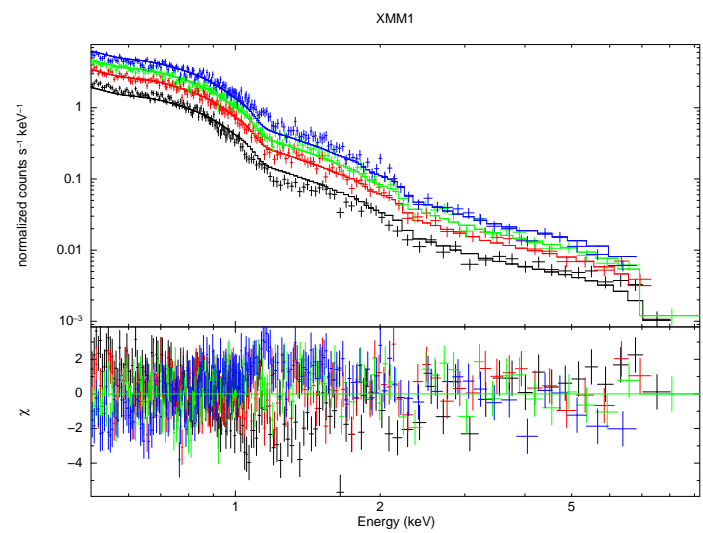

\begin{center}
\rotatebox{-90}{\FigureFile(70mm,90mm){alphafreeze2.eps}} 
\end{center}
\caption{
Simultaneous fitting result of the intensity-sliced spectra of XMM1 with the same covering fraction and variable normalizations
}
\label{alphafreeze1}
\end{figure*}

%IRAS lightcurve simulation

\addtocounter{figure}{0}
\begin{figure*}[ht] 
  \begin{minipage}{0.5\hsize}
    \begin{center}
     \rotatebox{-90}{ \FigureFile(60mm,80mm){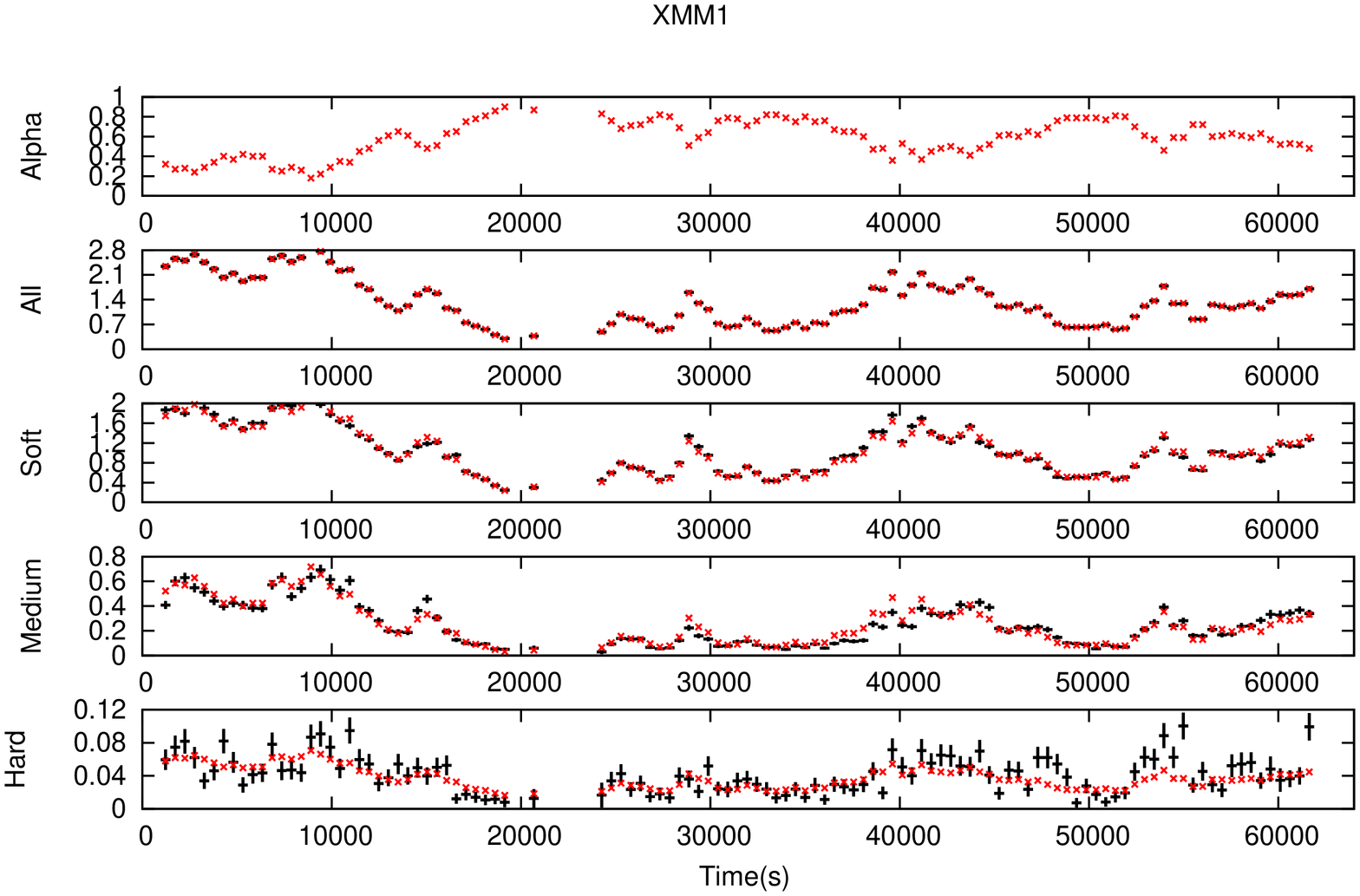}} \\
   \rotatebox{-90}{\FigureFile(60mm,80mm){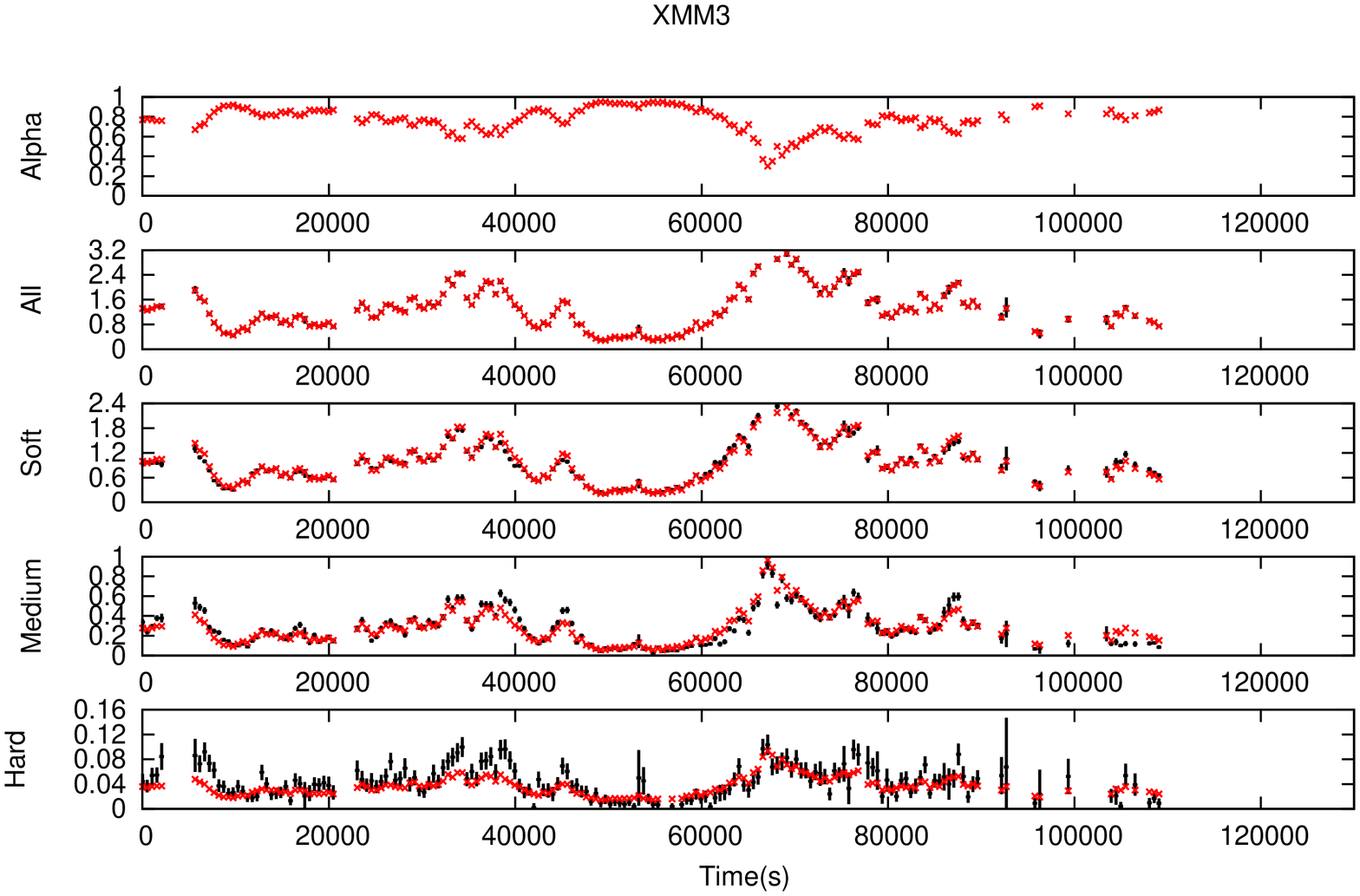}} \\
 \rotatebox{-90}{\FigureFile(60mm,80mm){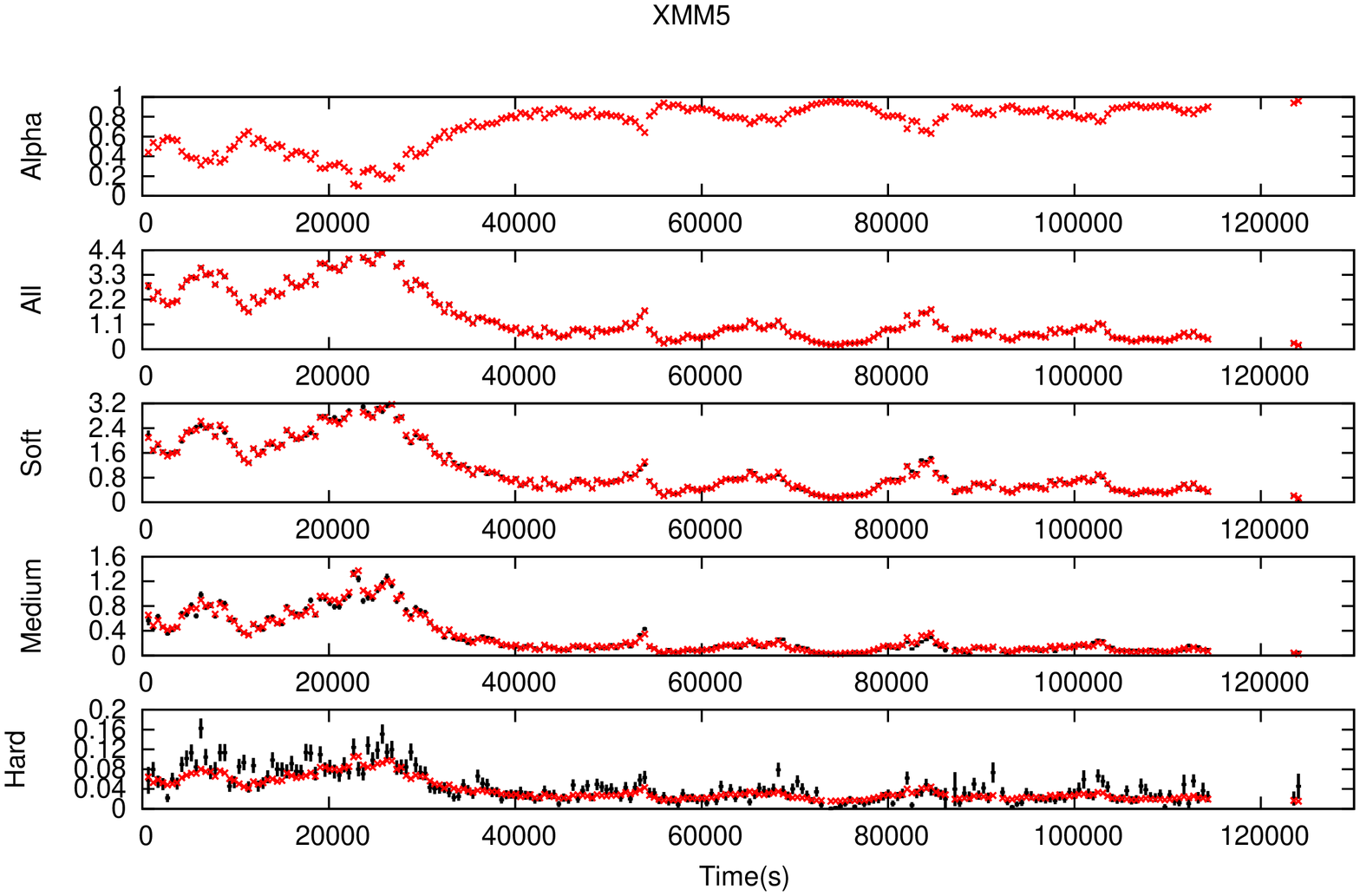}} \\
    \end{center}
  \end{minipage}
\begin{minipage}{0.5\hsize}
 \begin{center}
     \rotatebox{-90}{\FigureFile(60mm,80mm){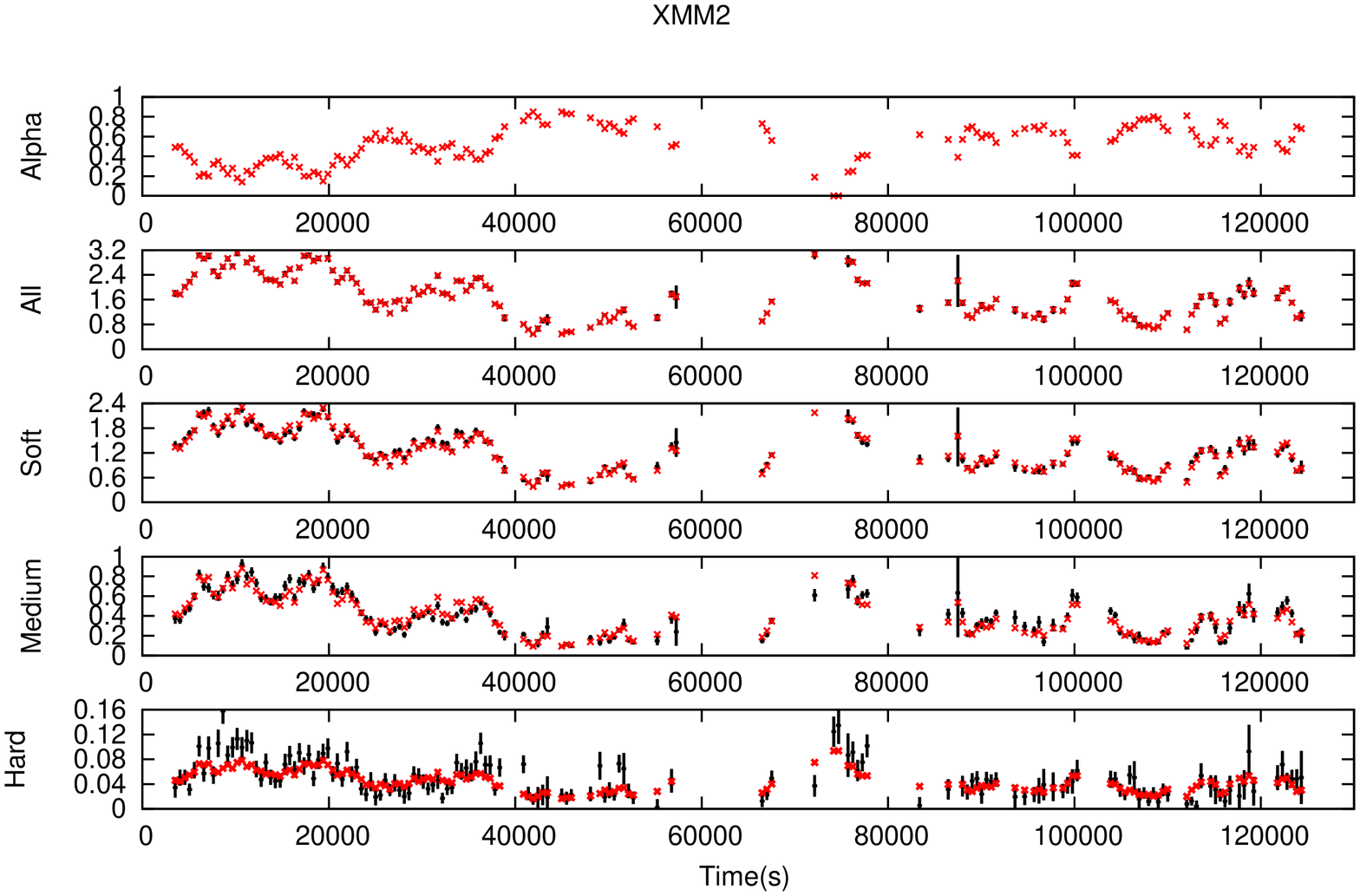}} \\
     \rotatebox{-90}{\FigureFile(60mm,80mm){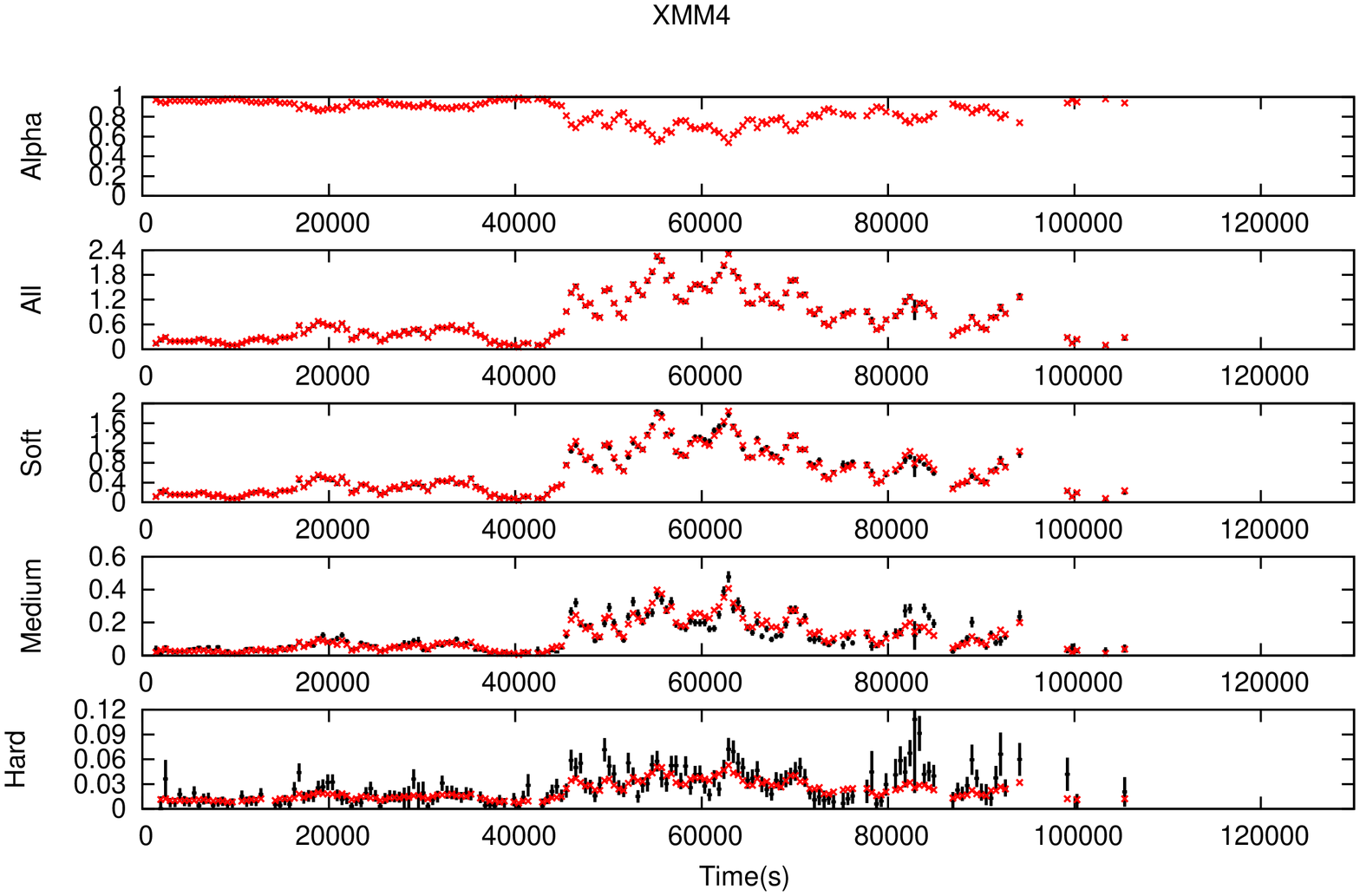}} \\
      \rotatebox{-90}{\FigureFile(60mm,80mm){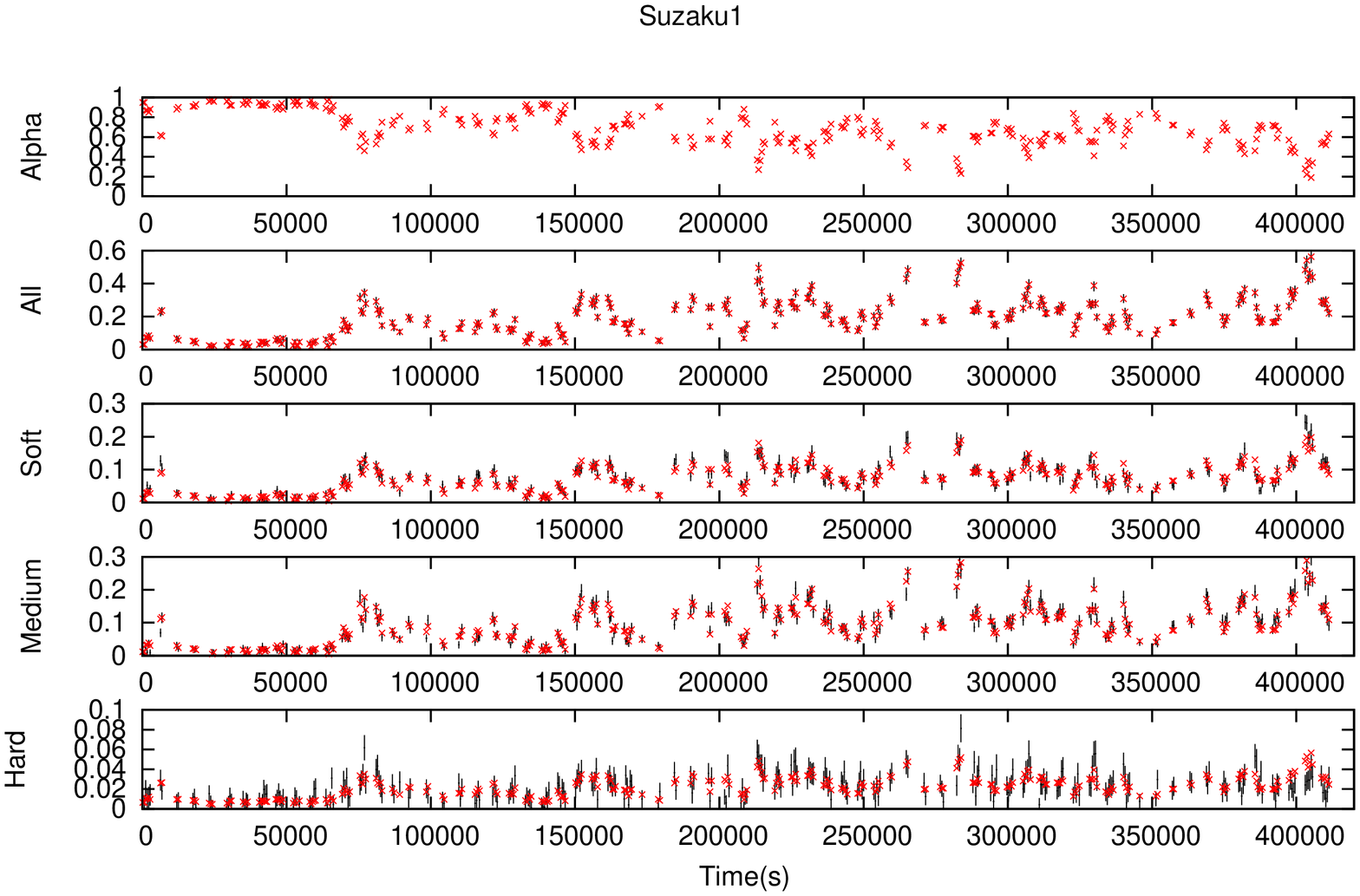}} \\
       \end{center}
  \end{minipage}
  \caption{Comparison of the observed light-curve and the model one of IRAS 13224$-$3809 for each observation sequence.
Top: Variation of the partial covering fraction calculated from the 0.5$-$10.0 keV light-curves assuming the VDPC model. 
Upper center/Center/Lower center/Bottom: The observed light-curve in 0.5$-$10.0~keV/0.5$-$1.0~keV/1.0$-$3.0~keV/3.0$-$10.0~keV (black) and the model one (red), respectively. 
Note that, in 0.5$-$10.0~keV, the black and red bins should agree in definition.}
\label{IRAS-fig4}
\end{figure*}
%%%%%%%%%%%%%%%%%%%%%%%%%%%%%%%%%%%%%%%%%%%%%%%%%%%%%%%%%%%%%%%%%%%%%%%%%%%%%%%%%%%%%

%%%%%%%%%%%%%%%%%%%%%%%%%%%%%%%%%%%%%%%%%%%%%%%%%%%%%%%%%%%%%%%%%%%%%%%%%%%%%%%%%%%%%

%IRAS rms spectra
\addtocounter{figure}{0}
\begin{figure*}[ht] 
  \begin{minipage}{0.5\hsize}
    \begin{center}
     \rotatebox{-90}{ \FigureFile(60mm,80mm){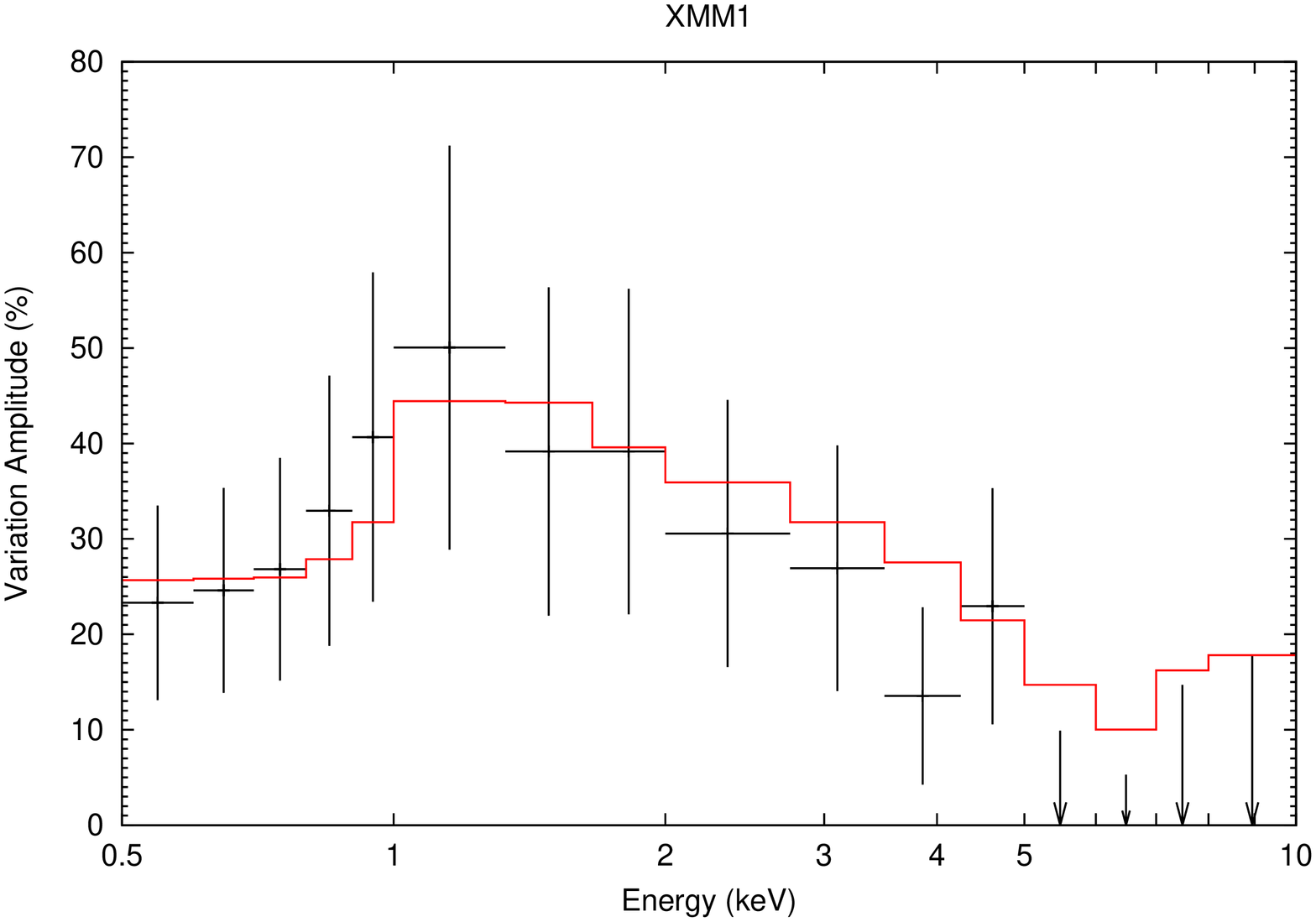}} \\
   \rotatebox{-90}{\FigureFile(60mm,80mm){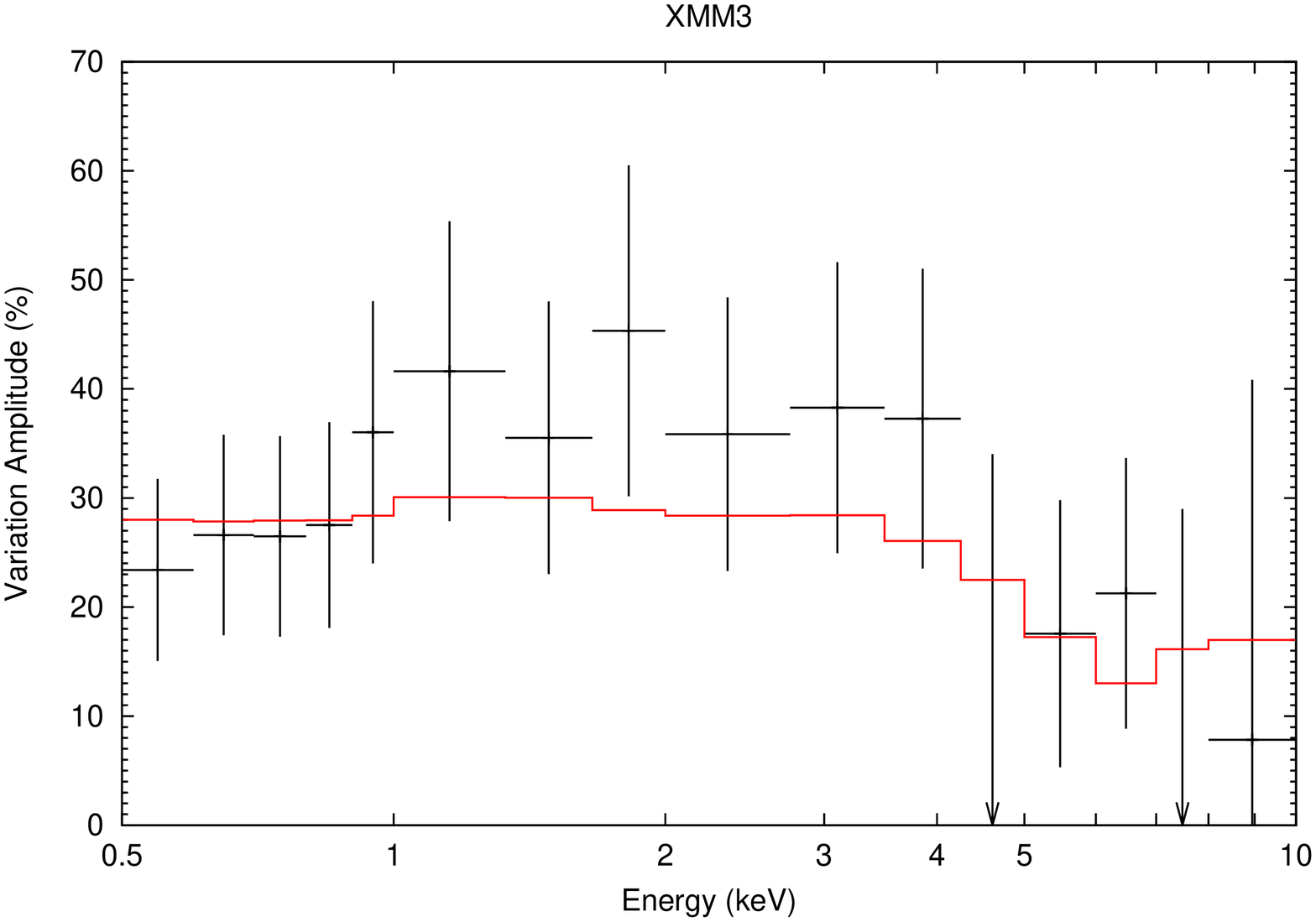}} \\
 \rotatebox{-90}{\FigureFile(60mm,80mm){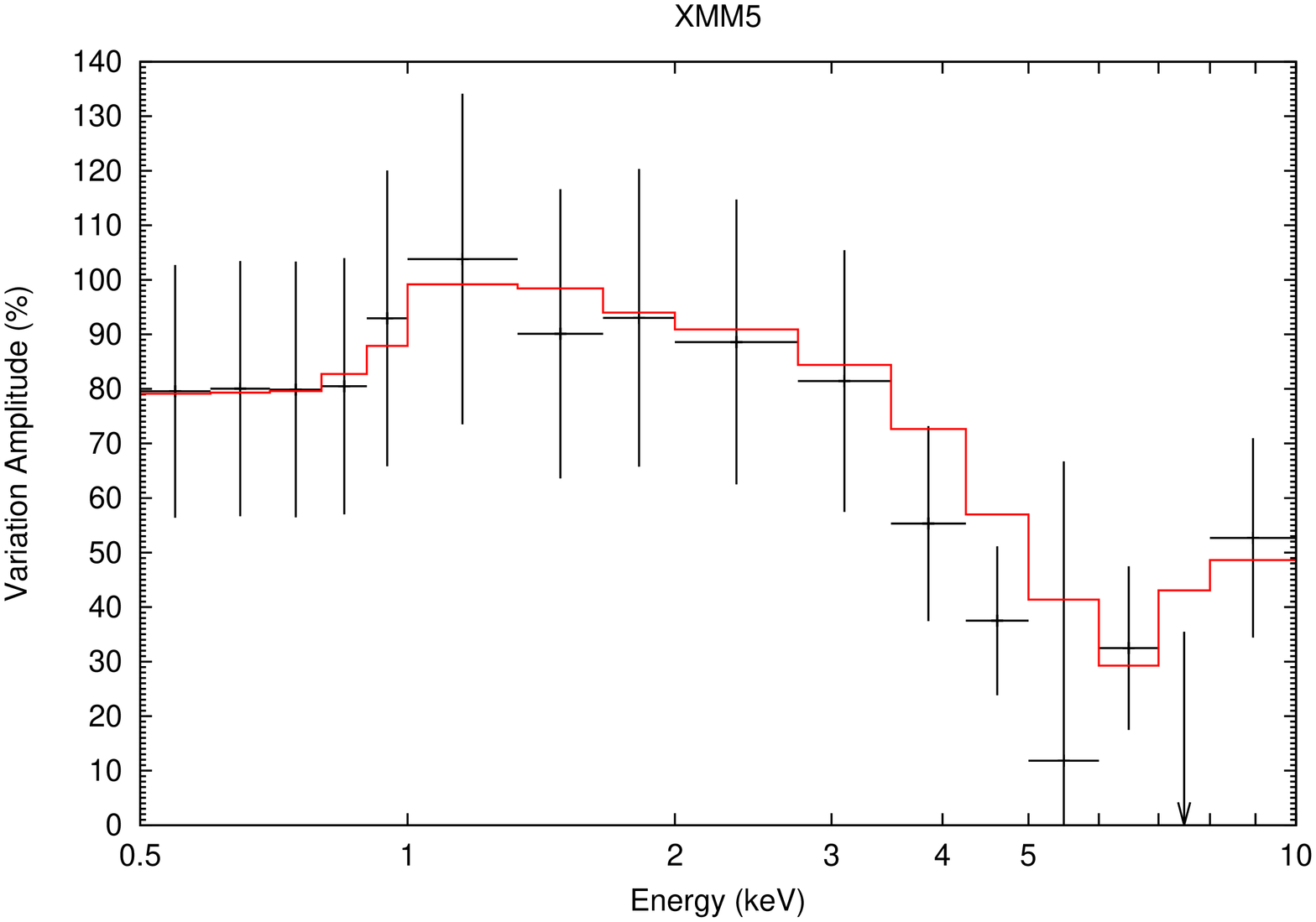}} \\
    \end{center}
  \end{minipage}
\begin{minipage}{0.5\hsize}
 \begin{center}
     \rotatebox{-90}{\FigureFile(60mm,80mm){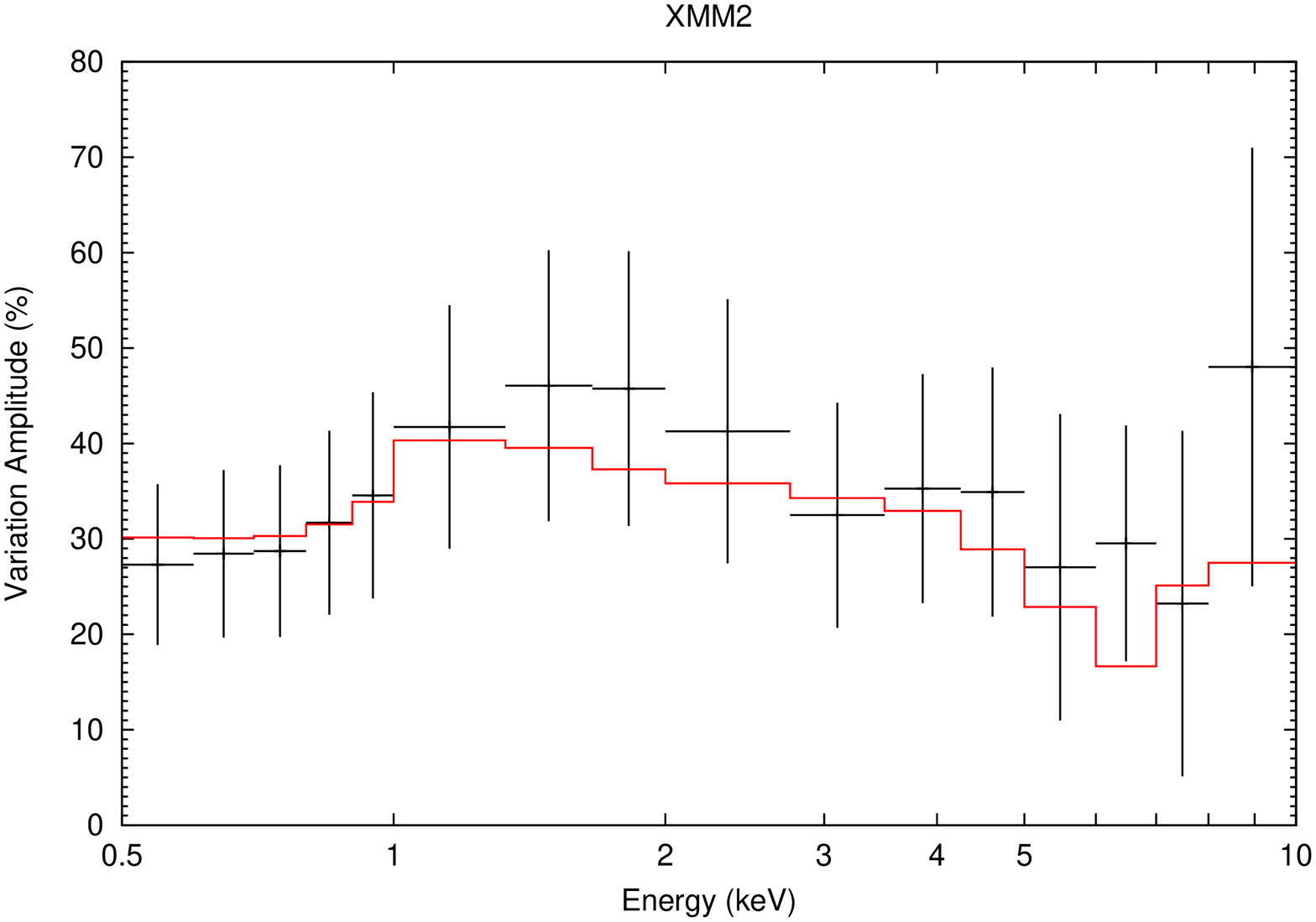}} \\
     \rotatebox{-90}{\FigureFile(60mm,80mm){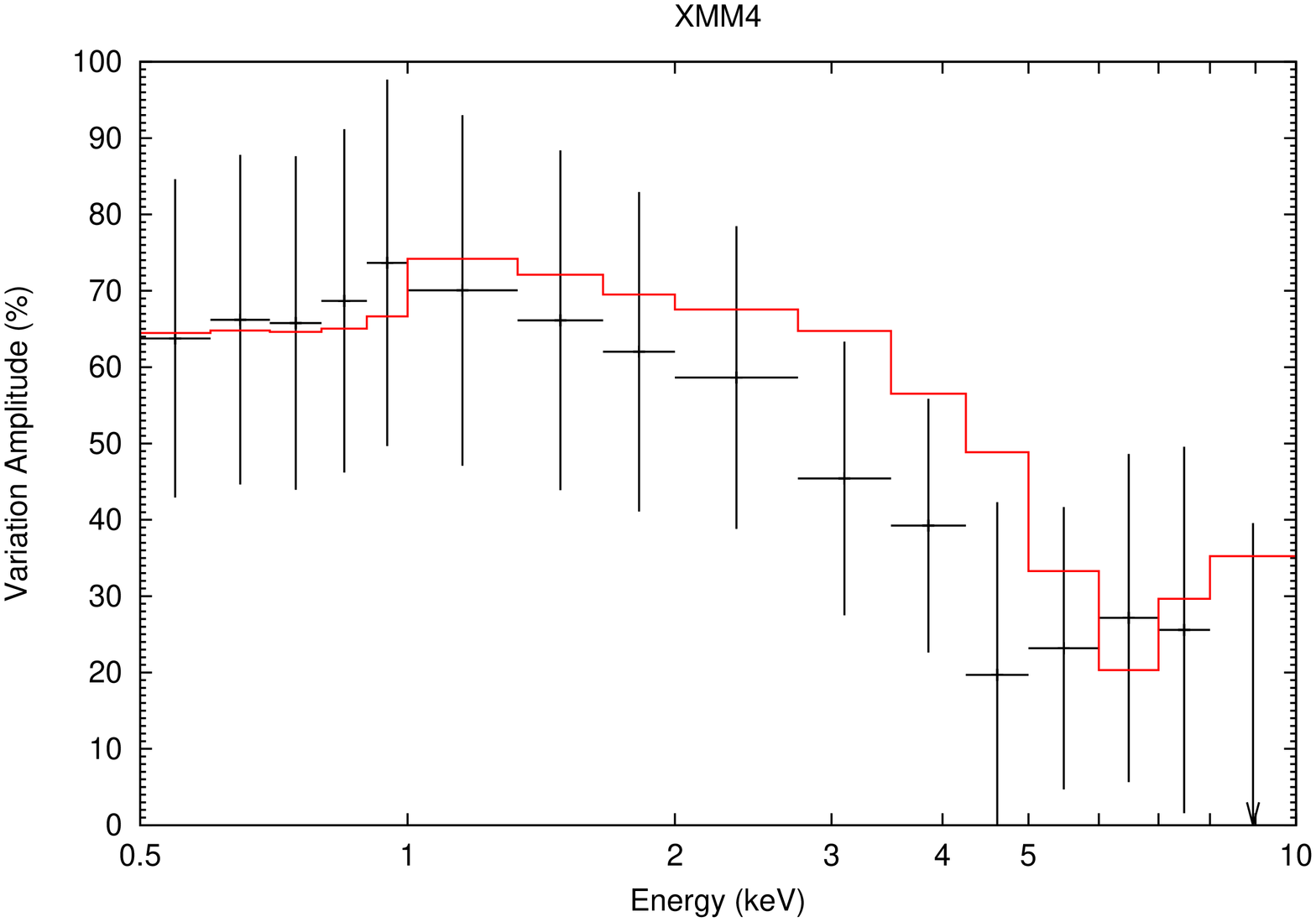}} \\
      \rotatebox{-90}{\FigureFile(60mm,80mm){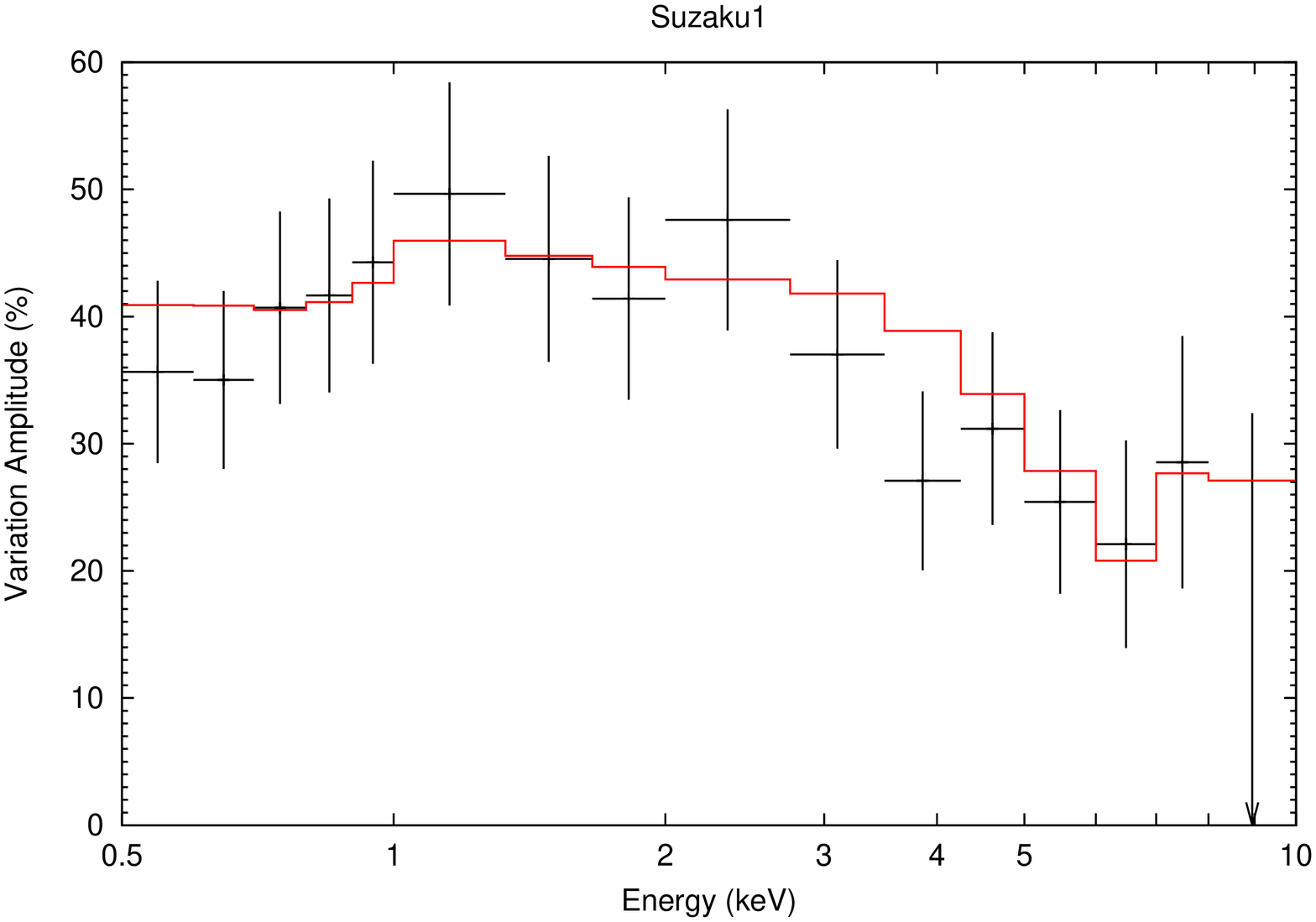}} \\
       \end{center}
  \end{minipage}
  \caption{RMS spectra for IRAS 13224$-$3809. Black points are calculated from the data, and red lines are calculated from the simulated  model light-curves.}
\label{IRAS_rms_spe}
\end{figure*}

%%%%%%%%%%%%%%%%%%%%%%%%%%%%%%%%%%%%%%%%%%%%%%%%%%%%%%%%%%%%%%%%%%%%%%%%%%%%%%%%%%%%%%

%1H 0707 RMS

\addtocounter{figure}{0}
\begin{figure*}[ht]
\begin{minipage}{0.5\hsize}
\begin{center}
       \rotatebox{-90}{ \FigureFile(60mm,80mm){1H_slice.eps}} 
       \rotatebox{-90}{\FigureFile(60mm,80mm){ark564_suzaku1.eps}} 
  \end{center}
\end{minipage} 
\begin{minipage}{0.5\hsize}
\begin{center}
       \rotatebox{-90}{ \FigureFile(60mm,80mm){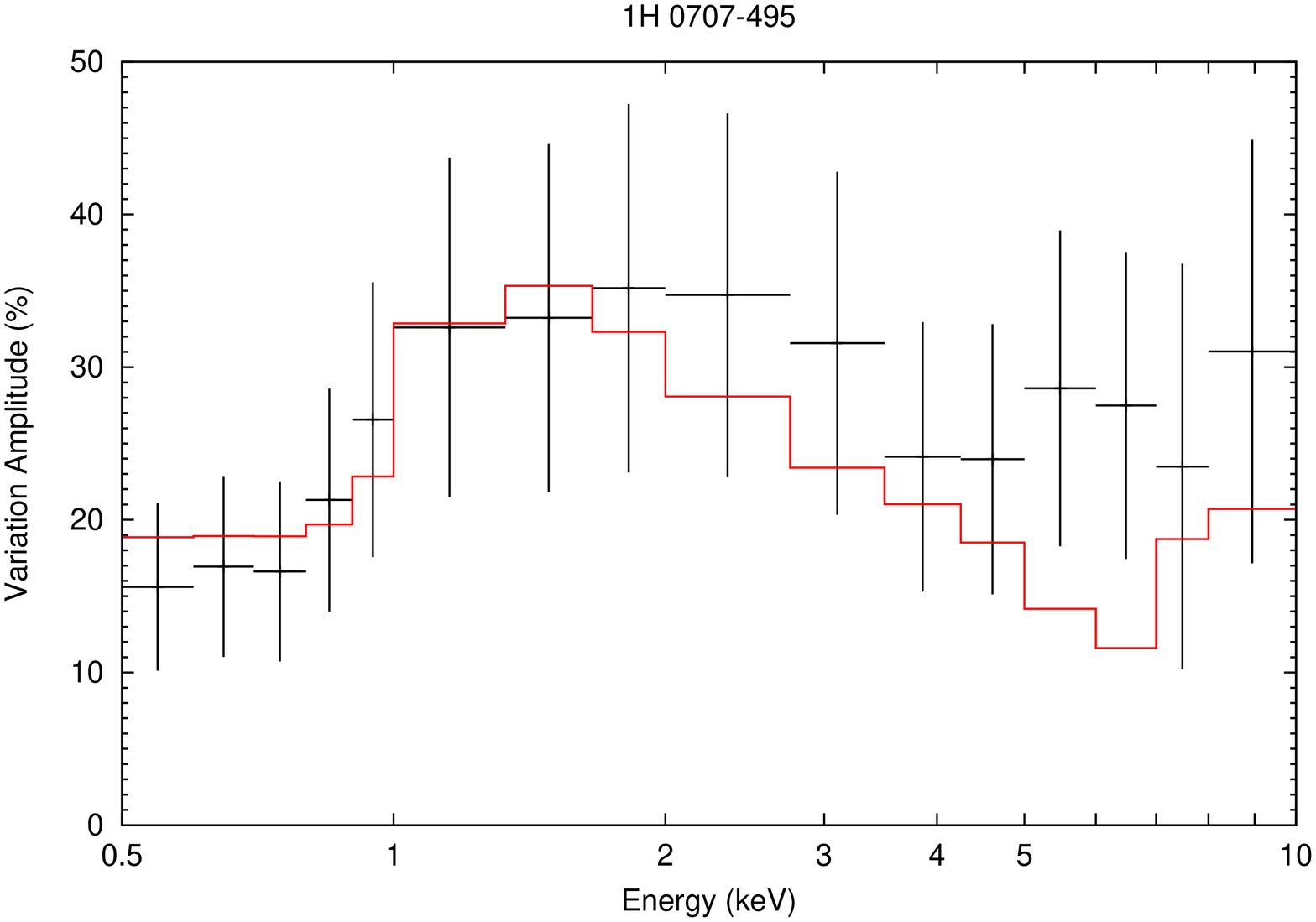}} 
       \rotatebox{-90}{\FigureFile(60mm,80mm){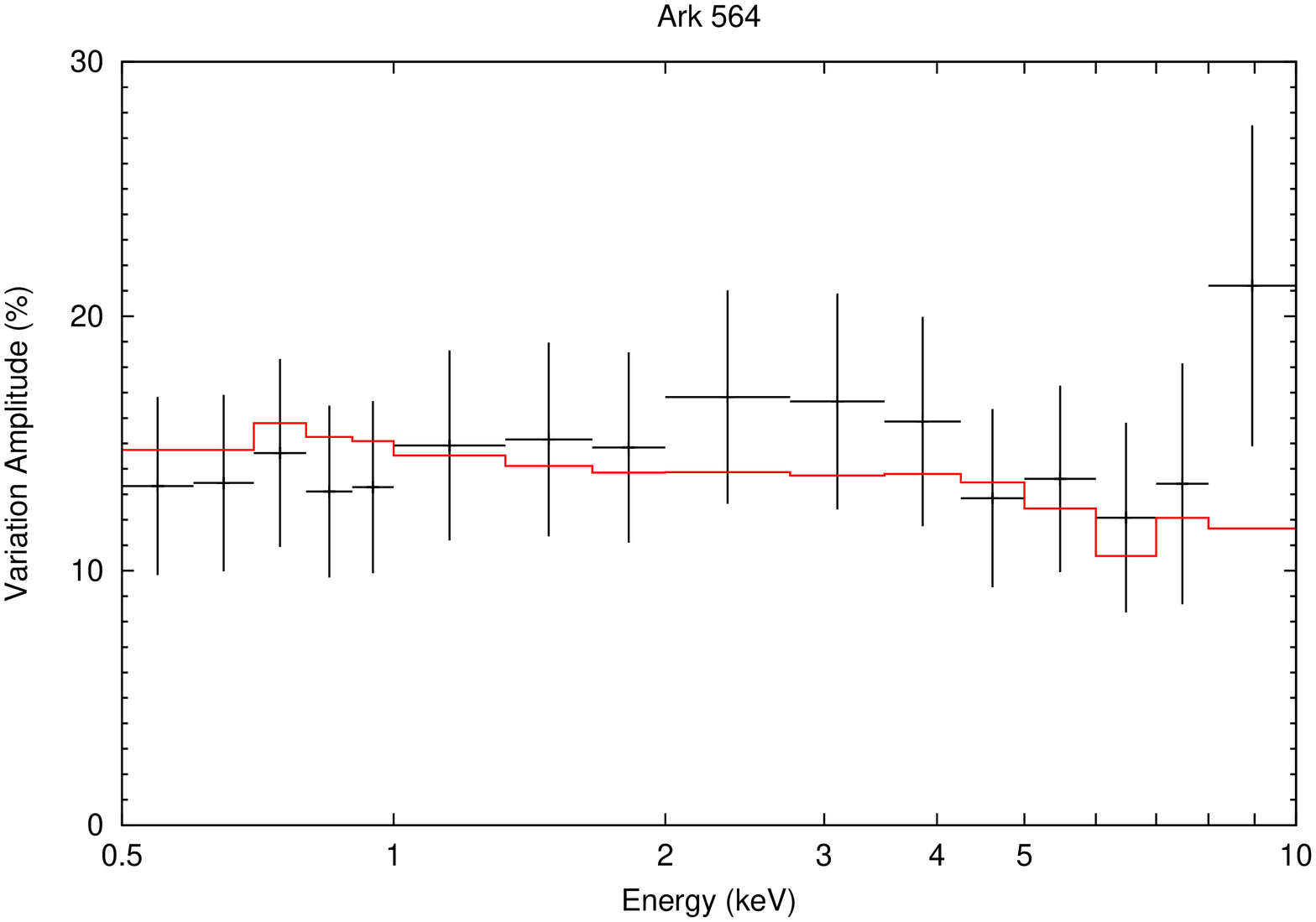}} 

  \end{center}
\end{minipage}  
  \caption{Intensity-sliced energy spectra (left) and the RMS spectrum (right) modeled with the VDPC model for the XMM-Newton PN data (ID: 0653510301) of 1H 0707$-$495 and the Suzaku data (ID: 702117010) of Ark 564. In the right, black points are calculated from the data, and red lines are calculated from the VDPC model.}
\label{1H0707}
\end{figure*}

%%%%%%%%%%%%%%%%%%%%%%%%%%%%%%%%%%%%%%%%%%%%%%%%%%%%%%%%%%%%%%%%

\addtocounter{figure}{0}
\begin{figure*}[ht]
\begin{center}
\rotatebox{-90}{\FigureFile(70mm,90mm){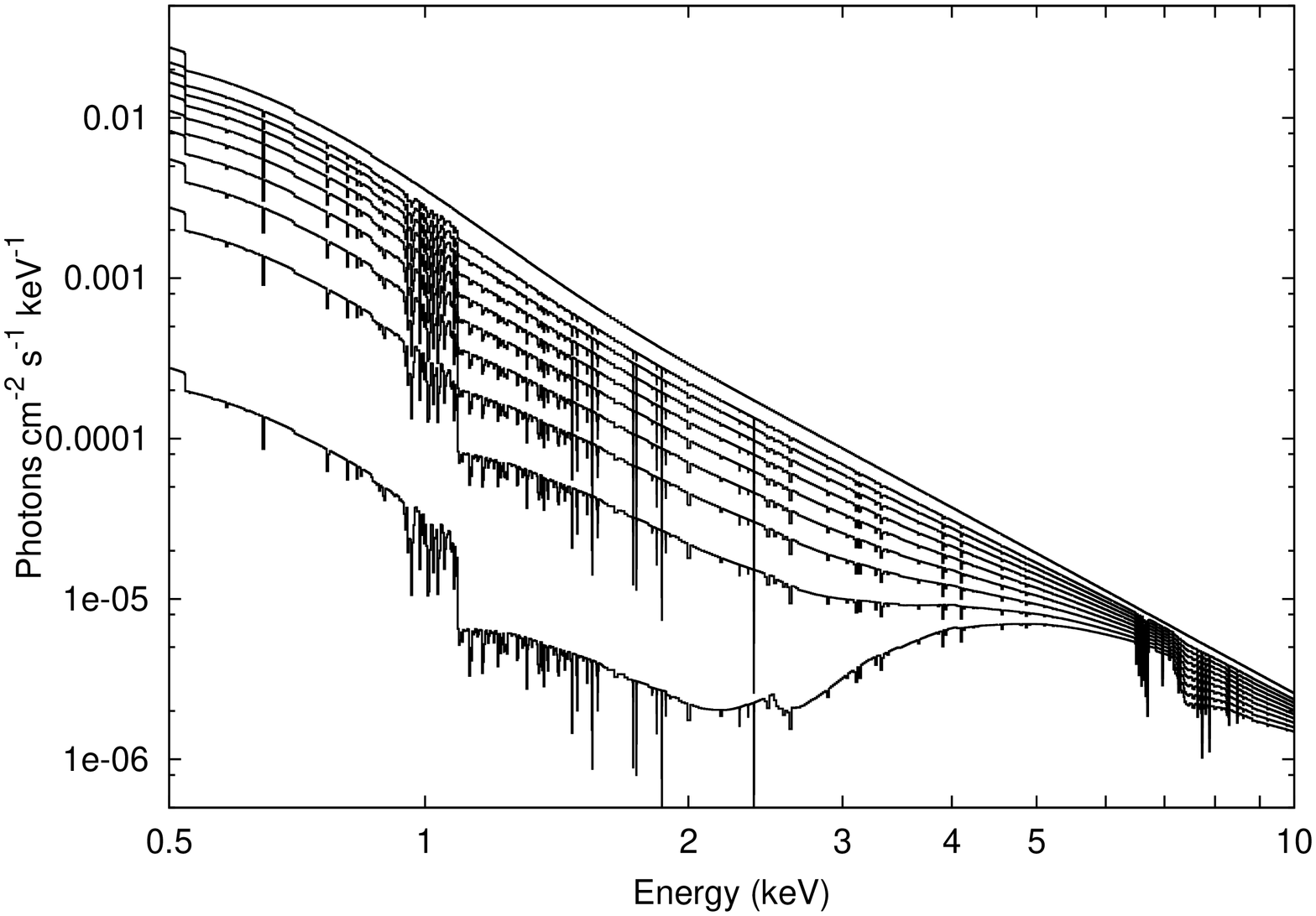}} \\
\rotatebox{-90}{\FigureFile(70mm,90mm){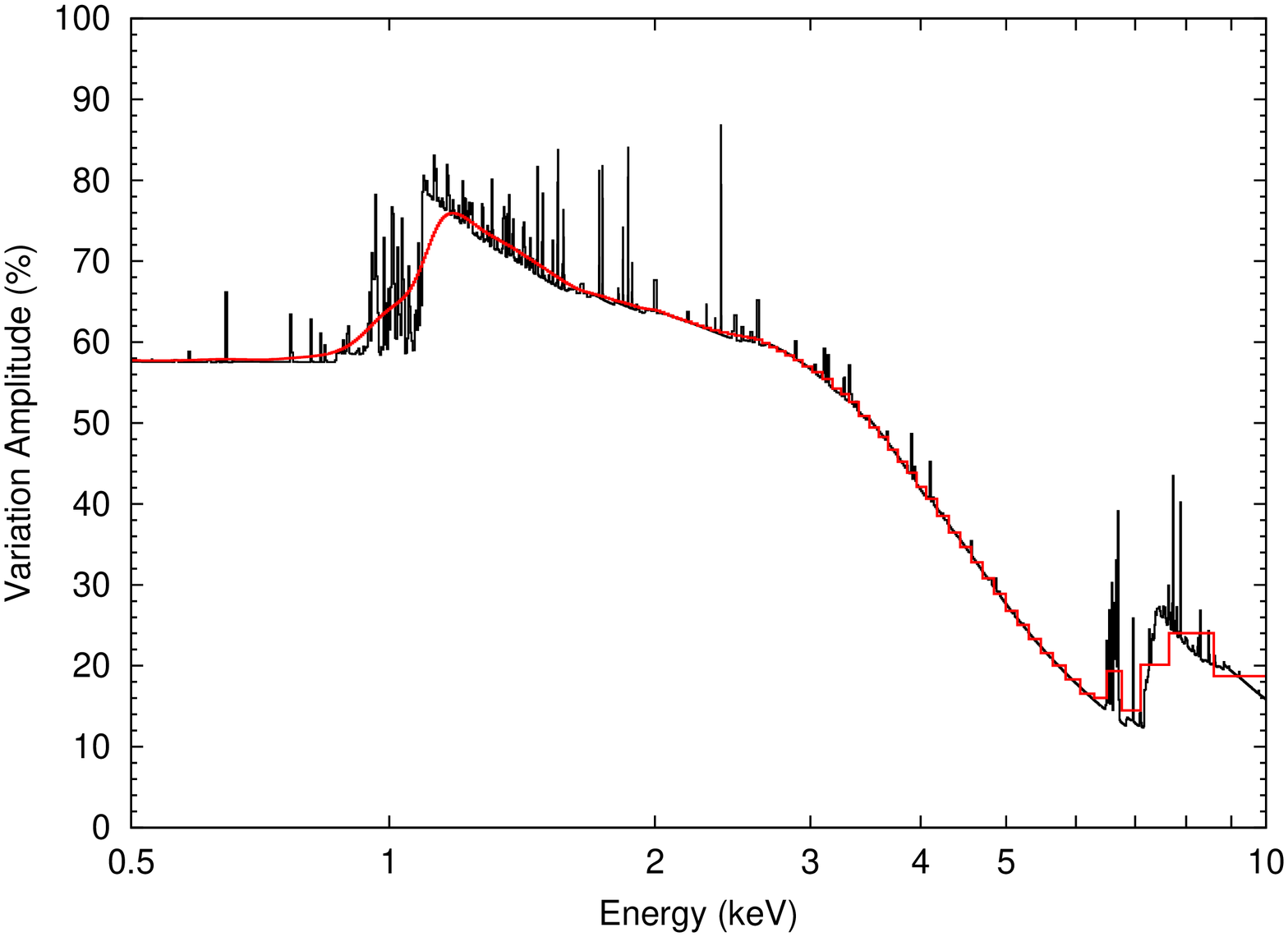}} \\
\end{center}
\caption{(Upper) VDPC model spectra where the covering fraction values are 0.01, 0.1, 0.2, 0.3, 0.4, 0.5, 0.6, 0.7, 0.8, 0.9, and 0.99 from top to bottom.
Other spectral parameters are taken from XMM1 (Table \ref{IRAS_para1}; except the 8.45~keV absorption line).
(Lower) Simulated RMS spectra when the covering fraction uniformly varies within 0.01--0.99. 
The black line is the one with an ultimate energy resolution ($\sim1$~eV),
whereas the red line assumes the XMM EPIC-pn response, with the same bin size of Figure \ref{fig1}.}
\label{RMSsimulation}
\end{figure*}

%%%%%%%%%%%%%%%%%%%%%%%%%%%%%%%%%%%%%%%%%%%%%%%%%%%%%%%%%%%%%%%%

\end{document}